# Experimental Facilities at the High Energy Frontier

*P. Jenni*
Albert-Ludwigs-University Freiburg, Germany, and CERN, Geneva, Switzerland

**Abstract**
The main theme of the lectures covered the experimental work at hadron colliders, with a clear focus on the Large Hadron Collider (LHC) and on the roadmap that led finally to the discovery of the Higgs boson. The lectures were not a systematic course on machine and detector technologies, but rather tried to give a physics-motivated overview of many experimental aspects that were all relevant for making the discovery. The actual lectures covered a much broader scope than what is documented here in this write-up. The successful concepts for the experiments at the LHC have benefitted from the experience gained with previous generations of detectors at lower-energy machines. The lectures included also an outlook to the future experimental programme at the LHC, with its machine and experiments upgrades, as well as a short discussion of possible facilities at the high energy frontier beyond LHC.

## 1 Introduction

Experimental facilities at the High Energy Frontier (HEF) is a very broad topic, impossible to cover in even four lectures. Already during the lectures, but even much more so in this write-up, the main theme will be the long journey to the Higgs boson discovery and beyond at the Large Hadron Collider (LHC).

The HEF is only one of the three main pillars of experimental particle physics. It complements experiments that study the physics at the intensity frontier and at the cosmic frontier. The three areas have considerable overlap that provide for exciting synergies. Without exaggeration one can claim that particle physics has never been such an exciting research field with great promises for new fundamental discoveries as we live through these years. The HEF in itself is a broad and lively field with complementary approaches using either proton-proton (pp), electron-positron (e+e-) or eventually electron-proton (ep) collisions. For these lectures, however, the LHC will be in the focus, given the topical discovery of the Higgs boson and LHC's unique status of being the only running high energy collider for many years to come.

The discovery of a scalar boson, which shows within the present statistical precision achieved all the expected properties for the famous Higgs boson, announced by the ATLAS and CMS Collaborations on 4[th] July 2012 [1,2], was the culminating experimental triumph for the Standard Model [3-11]. The Standard Model (SM) of particle physics is one of the most remarkable achievements of physics over the past 50 years. Its descriptive and predictive power has been experimentally demonstrated with unprecedented accuracy in many generations of experiments ranging from low to high energies. The SM comprises the fundamental building blocks of all visible matter, with the three fermion families of quarks and leptons, and their interactions via three out of the four fundamental interactions mediated by bosons, namely the massless photon for the electromagnetism, the heavy W and Z bosons for the weak force, these two interactions unified in the electroweak theory, and the massless gluons for the strong interaction.

In order to solve the mystery of generation of mass, a spontaneous symmetry-breaking mechanism was proposed introducing a complex scalar field that permeates the entire Universe. This mechanism, known as Brout-Englert-Higgs (BEH) mechanism [6-9], gives the W and Z their large masses and leaves the photon massless. Interaction with the scalar field generates masses to the quarks and leptons in proportion to the strength of their couplings to it. This field leads to an additional massive scalar boson as its quantum, called the Higgs boson. After the discovery of the W and Z bosons in the early 1980s, the hunt for the Higgs boson, considered to be the keystone of the SM, became a central theme in particle physics, and also a primary motivation for the LHC. Finding the Higgs boson would establish the existence of the postulated BEH field, and thereby marking a crucial step in the understanding of Nature.

There is a vast body of literature available for 'telling the story' of the Higgs discovery at the LHC. Somehow unjustified and selectively, the following accounts are heavily based on a few didactic articles co-authored by the lecturer with longtime colleagues since the early years of the LHC adventure [12,13] and hadron collider experiments even before [14]. For scientifically rigorous accounts and full references the reader is referred to the original publications [1,2] and the full references therein, as well as to the updated public reference lists of publications by ATLAS [15] and CMS [16].

## 2  Hadron Collider Experiments Preceding the LHC

The history of hadron colliders started almost 45 years ago. A dedicated account of the evolution of hadron collider experimentation is given in [14], containing also detailed references, where the very impressive growth and sophistication, both for the detector concepts as well as for the analysis methods, is discussed.

The first hadron collider was the about 1 km circumference CERN Intersecting Storage Rings (ISR), commissioned in 1971 with proton-proton (pp) collision energies E between 23 and 63 GeV. Experiments were located at eight beam crossing points of the two separate magnet rings. The CERN SPS Collider (SppbarS), colliding p and antiprotons (pbar) with E = 546 and 630 GeV, followed in 1981. The pbars were produced using intense proton beams hitting a target, and collected in a dedicated accumulator ring, and their phase space was reduced to dense bunches by an electromagnetic feedback system derived from sensing energy fluctuations in the beam itself (stochastic cooling). The 1.3 Tesla conventional bending magnets were housed in the 6.9 km SPS tunnel. The two general-purpose detectors, UA1 and UA2, where the W and Z were discovered, shared the collider with a few smaller experiments. The Fermilab ppbar Tevatron Collider with E = 2 TeV used 4.2 Tesla superconducting magnets in a 6.3 km ring. It began operation in 1987, serving two major detectors, CDF (Collider Detector Facility) and DØ (named after its location on the ring), which provided the top quark discovery, and also a few small specialized experiments. Finally, the pp Superconducting Super Collider (SSC) in Texas would have had E = 40 TeV, but was cancelled by the U.S. Congress in 1993 well before its construction completion.

Hadron collider detectors have evolved in size and complexity as physics questions have changed, and as the technology advanced. Each generation of detectors built on the previous experience. The growth in size stemmed in part from the increase in typical particle energies, but also in response to the need for higher precision and hence more sophisticated detectors.

The early ISR experiments, as customary from previous fixed target experiments, had rather specific physics goals that were addressed with detectors covering limited solid angles using the detector technologies from the fixed target era. Toward the end of the ISR period, new discoveries stimulated a new detector paradigm. The requirement to study new massive particles, such as the W and Z that transmit the weak force, and the search for the top quark, in which several decay particles emerge over a large angular region and at large momenta transverse to the beam axis ($p_T$), stimulated experiments to become more inclusive in their coverage. Thus, starting with the UA1 and UA2 experiments, the general purpose detectors followed a more hermetic design with cylindrical shells of subdetectors surrounding the beam vacuum pipe and covering nearly the full 4π solid angle. The innermost 'tracker' layer records the ionization tracks of charged particles moving in a magnetic field to allow the measurement of their momenta, followed by 'calorimeters' to measure the energies of hadrons, electrons and photons, and an outer layer that measures and identifies the muons which penetrate the calorimeter. Implementation in specific experiments varied, depending on the physics emphasis and the available technology. Figure 1 illustrates with some examples this detector evolution.

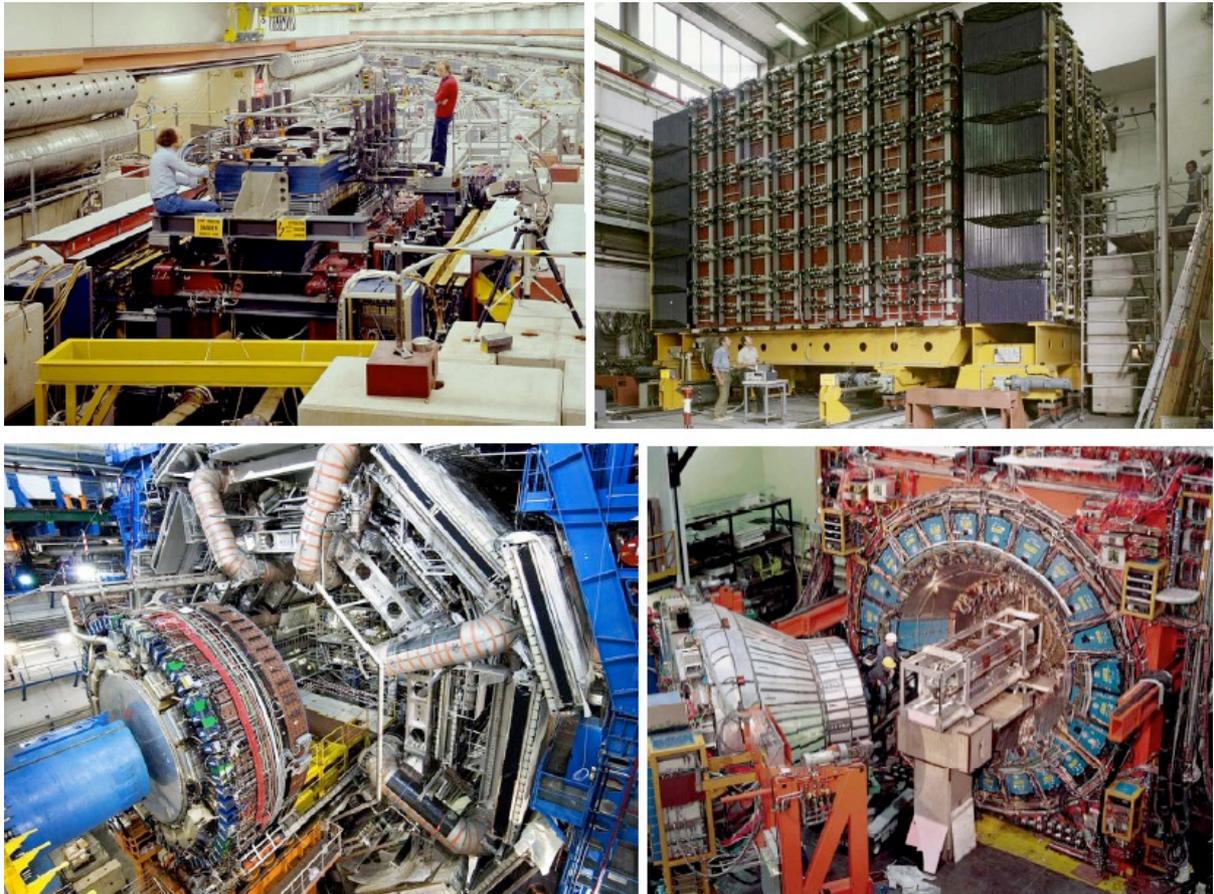

**Fig. 1:** Clockwise from top left the R702 experiment at the ISR, the UA1 experiment at the SppbarS, the CDF experiment at the Tevatron, and the ATLAS experiment at the LHC, whose approximate diameters are 6, 10, 12 and 24 m respectively.

The trackers evolved from multiwire proportional chambers and drift chambers that measured the location of charged particle ionizations with precisions of hundreds of μm, to higher resolution scintillating fibers and ultimately silicon microstrip and pixel detectors with few μm resolutions. To allow track momentum measurement, solenoids with magnetic field along the beam axis were favoured as this optimized the transverse momentum resolution needed by the high $p_T$ physics programme. The fields increased from about 1 Tesla in earlier detectors to 4 Tesla in the LHC CMS solenoid. The resolution improvement from silicon detectors brought the new capability to sense the short distance between the production and decay of hadrons containing a b-quark, a signature of many kinds of new physics.

The calorimeters were designed to fully contain the energy deposits of electrons, photons and hadrons through a cascade of collisions, each producing several lower energy particles until the shower multiplication dies. The shower containment distance is a function of the radiation length for electrons and photons and of the nuclear interaction length for hadrons. In sampling calorimeters, as used in all hadron collider experiments before the LHC, layers of absorber are interleaved with active detectors that measure the ionization signal which is proportional to the incident particle energy. The two length scales dictate different design choices for the electromagnetic (em) and hadronic (had) calorimeter sections. The energy resolution of the calorimeters is controlled by statistical fluctuations in the shower process and in the number of particles traversing the active layer, as well as by calibration variations and inhomogeneity due to, for example, cables and supports. The $4\pi$ calorimeter coverage allowed measurement of almost all produced particles so as to determine the total $p_T$ of visible particles. Since the initial colliding hadrons have zero $p_T$, large 'missing' transverse energy ($E_T^{miss}$) can be ascribed to non-interacting particles such as neutrinos.

Muons, the only observable charged particles that can penetrate the calorimeter and magnet, are identified in large-area position detectors in the outer detector shell. The muon detectors were typically embedded in the iron return yokes of the magnet and relied on a precise momentum measurement in the tracker, but in the LHC ATLAS detector with its open air-core toroids outside the calorimeter a second precise momentum measurement can be made.

With the increased luminosity, the total collision rate grew from $3 \times 10^5$ Hz at the SppbarS, to $3 \times 10^7$ Hz at the Tevatron and now to $10^9$ Hz at the LHC. Although the ability to record data for offline analysis has grown with modern computing technology, the number of bytes of data describing an event has also grown and it remained until recently impractical to record collisions at more than a few 100 Hz. Complex 'trigger' systems have evolved to select the few potentially interesting events from the flood of all collisions. Reliable triggering is key for the success of hadron collider experiments, as a rejected event is lost forever. The SppbarS triggers used fast indications of single particles like electrons and muons to signal an interesting event, although a smaller stream was selected by more elaborate analyses in microprocessors. At the Tevatron, three-level triggers were introduced in which the first deadtime-less decisions were made in single chip processors based on tracker, calorimeter or muon information, followed by microprocessor decisions combining several subdetectors, and finally a farm of processors running a simplified version of offline processing. The multiple level triggering gave data logging at 50 Hz with a dead time loss of about 5%. The LHC experiments retained the multi-level triggering with much more power at each stage to achieve fully efficient logging rates of a few 100 Hz. The growing luminosity also gives an increase in the number of events besides the one of interest occurring in the same bunch crossing. At the LHC, such 'pileup' gives dozens of extra events superimposed on the one of interest, and is a major challenge for physics analyses.

## 3   The LHC Project

The LHC project must be seen as a global scientific adventure, combining the accelerator complex, the experiment collaborations with their detectors, a worldwide computing grid, and a motivating theory community, that started more than 30 years ago. Obviously the main issue is the fundamental physics it addresses, but the project itself is truly 'a marvel of technology', a detailed account of which can be found in [17], and specific technological highlights are featured in [18], from which some of the technical discussion is reproduced here.

### 3.1   Historical Setting and Time Line

A comprehensive history of the years leading to the LHC can be found in [19]. The great success in making the experimentally "clean" W and Z boson discoveries, despite the huge hadronic backgrounds, at the CERN SppbarS Collider in the early 1980s was crucial for the community to dare to even dream of a future powerful high-energy hadron collider in order to make a decisive search for the Higgs boson. The idea that the tunnel for the, at that time, still future Large Electron-Positron (LEP) machine should be able to house, at some distant time, the LHC, was already in the air in the late 1970s [20]. Thankfully, those leading CERN at the time had the vision to plan for a tunnel with dimensions that could accommodate it. Enthusiasm for an LHC surfaced strongly in 1984 at a CERN-ECFA workshop in Lausanne entitled "LHC in the LEP Tunnel", which brought together working groups that comprised machine experts, theorists and experimentalists.

With the promise of great physics at the LHC, several motivating workshops and conferences followed, where the formidable experimental challenges started to appear manageable, provided that enough R&D work on detectors would be carried out. Highlights of these "LHC experiment preliminaries" were the 1987 Workshop in La Thuile of the so-called "Rubbia Long-Range Planning Committee" and the large Aachen ECFA LHC Workshop in 1990. Finally, in March 1992, the famous conference entitled "Towards the LHC Experimental Programme", took place in Evian-les-Bains, where several proto-collaborations presented their designs in "Expressions of Interest". This was also the time when CERN created an international peer review committee for the LHC experiments (LHCC). Moreover, from the early 1990s, CERN's LHC Detector R&D Committee (DRDC), which reviewed and steered R&D collaborations, greatly stimulated innovative developments in detector technology.

It cannot be stressed enough how important the many years of R&D were that preceded the final detector construction for the LHC experiments. Technologies had to be taken far beyond their state-of-the-art of the late 1980s in terms of performance criteria in the anticipated harsh LHC environment, like granularity and speed of readout, radiation resistance, reliability, but also considering buildable sizes of the detector components and number of units, and very importantly at an affordable cost. For many detector subsystems there were initially a few parallel developments pursued as options, because it was not guaranteed from the onset that a given proposed technology would finally fulfil all the necessary requirements. Increasingly more realistic prototypes were developed, in a learning process for both the detector communities and the industries involved.

Some of the major technology decisions were taken by the Collaborations before the submission of the Technical Proposals to the LHCC end of 1994, which were finally approved early in 1996. For other choices the R&D needed more time, and they could only be made in the subsequent years from 1996 to the early 2000s, thereby defining the timing for the final Technical Design Reports of the various detector components.

The long duration of the LHC project until the Higgs discovery is illustrated in Table 1, with a few selected milestones concerning the LHC and the general-purpose experiments.

**3.2   The Collider**

In the LHC the two counter-rotating beams travel in separate channels in the arcs around the ring and consist of many particle bunches separated by a small distance. They are made to collide at the centre of the detectors with a small crossing angle. In the case of the LHC, the nominal number of bunches is 2808, the distance between bunches is 7.5 m and the crossing angle 285 µrad. The layout of the collider is shown in Fig.2. There are eight arcs (bending radius of 2804 m), where the beams are bent, and eight straight-sections used for utilities or collision insertions. Four insertions are equipped with experimental detectors, where the two counter-rotating beams can be brought to collision.

**Table 1:** The LHC Timeline

| | |
|---|---|
| 1984 | Workshop on a Large Hadron Collider in the LEP tunnel, Lausanne, Switzerland. |
| 1987 | Workshop on the Physics at Future Accelerators, La Thuile, Italy. The Rubbia "Long-Range Planning Committee" recommends the Large Hadron Collider as the right choice for CERN's future. |
| 1990 | LHC Workshop, Aachen, Germany (discussion of physics, technologies and detector design concepts). |
| 1992 | General Meeting on LHC Physics and Detectors, Evian-les-Bains, France (with four general-purpose experiment designs presented). |
| 1993 | Three Letters of Intent evaluated by the CERN peer review committee LHCC. ATLAS and CMS selected to proceed to a detailed technical proposal. |
| 1994 | The LHC accelerator approved for construction, initially in two stages. |
| 1996 | ATLAS and CMS Technical Proposals approved. |
| 1997 | Formal approval for ATLAS and CMS to move to construction (materials cost ceiling of 475 MCHF). |
| 1997 | Construction commences (after approval of detailed Technical Design Reports of detector subsystems). |
| 2000 | Assembly of experiments commences, LEP accelerator is closed down to make way for the LHC. |
| 2008 | LHC experiments ready for pp collisions. LHC starts operation. An incident stops LHC operation. |
| 2009 | LHC restarts operation, pp collisions recorded by LHC detectors. |
| 2010 | LHC collides protons at high energy (centre of mass energy of 7 TeV). |
| 2012 | LHC operates at 8 TeV: discovery of a Higgs boson. |

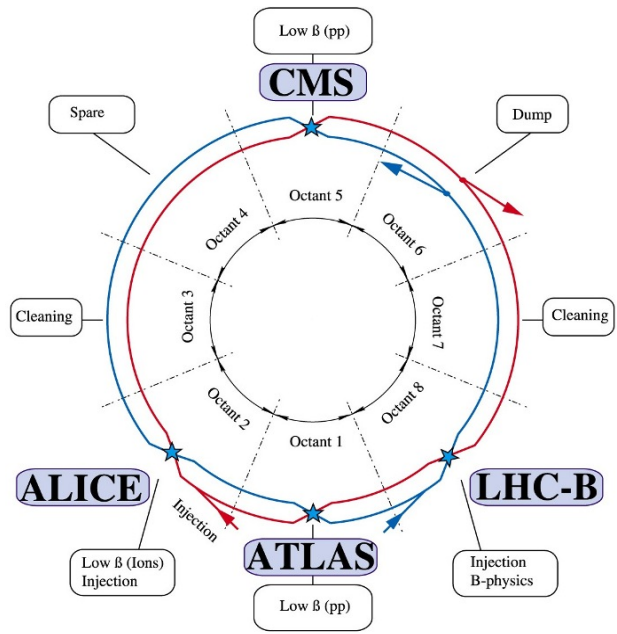

**Fig. 2:** Layout of the Large Hadron Collider

Just to recall, the two most important parameters of a proton-proton collider are the energy of the collisions and the luminosity, a parameter proportional to the number of collision events per second. The energy is related to the discovery potential of new particles of higher mass and the luminosity to the production of a relevant number of the desired events in a reasonable running time. The luminosity is a measure of the quality of the beams and their collisions, which is the result of careful design and mastering of several phenomena.

The **beam energy** is proportional to the product of B (the magnetic field of the main dipole magnets) and ρ (the bending radius of the arcs). Obviously, since ρ was fixed by the dimensions of the LEP tunnel, it was important to aim at the highest possible field B. Prior to the LHC, three large accelerators are/were based on superconducting magnets: the Tevatron (Fermilab, Chicago), HERA (DESY, Hamburg) and RHIC (BNL, Brookhaven). All of these make/made use of classical NbTi cables cooled by helium at a temperature slightly above 4.2 K (He-I). In each case the fields are below or around 5 Tesla. The choice for LHC was to use NbTi superconductors cooled at a lower temperature, namely 1.9 K provided by superfluid helium (He-II). The gain due to the lower temperature is about 1.5 Tesla or an additional 1.3 TeV in energy. It is interesting to note that the solution adopted for the LHC reconciled two requirements, namely the quest for the highest possible magnetic field and a design such that the magnets could be constructed in existing industry. This design with 8.3 Tesla magnetic field in the dipoles resulted in a nominal beam energy of 7 TeV. Out of the almost ten thousands magnets forming the collider, the 1232 main dipole magnets (length 16 m, mass 27 tons) represented a major design and constructional effort.

The solution of NbTi conductors at 1.9 K offers another important advantage at the expense of a more complex cryogenic system. It is due to the peculiar transport properties of pressurized superfluid helium, such as high heat capacity, low viscosity, and good effective thermal conductivity: it permits to keep the very large total helium mass static by cooling the long string of magnets with only a very small flow of liquid. The total cold mass of about 35000 tonnes operating at 1.9 K is the coldest spot in the Universe.

The considerable beam power (362 MJ) and the large electromagnetic energy stored in the magnets (11 GJ for the complete system) require a very sophisticated protection system to prevent damage in case of beam guiding problems or resistive transition (quench). In fact each of the two beams have an energy sufficient to melt 500 kg of copper and the electromagnetic energy of the magnets, if not properly discharged, can provoke large damages.

Another important element is the beam vacuum. The requirements for the beam vacuum are imposed by the beam lifetime and the background to the experiments. The interactions between the protons and rest gas are driven by two processes, i.e. single proton-nuclear collisions in which a proton is lost, and multiple small-angle Coulomb collisions, which provoke an increase of the size of the beam and a decrease of the luminosity. To insure a beam lifetime of a few hours, the residual pressure in the vacuum chamber should not exceed $10^{-10} - 10^{-11}$ mbar. The vacuum chamber in the LHC consists basically two types: i) chambers made from stainless steel at 1.9 K in the arcs; ii) chambers made from copper at ambient temperature in the straight sections covering about 6 km of the circumference.

The *luminosity L* is related to the properties of the colliding beams at the collision point and it is measured in $cm^{-2} s^{-1}$. It is proportional to the square of number of particle per bunch (N), the number of bunches around the ring ($n_b$) and inversely proportional to the transverse dimensions of the beams at the collision point. A small correction factor F takes into account the small angle of the beams at crossing. The product of L with the cross-section σ of the process to be investigated gives the average number of events produced per second. L varies with time, since the stored beams degrade during the collision run. The integrated luminosity is the integral L(t)dt over a certain time period. The luminosity can be considered as a figure of merit of the global quality of the machine. During the 2012 run, with beam energy of 4 TeV, the highest initial luminosity was $7.7 \cdot 10^{33}$ $cm^{-2} s^{-1}$. The integrated luminosity collected by ATLAS and CMS in 2012 was 21.5 $fb^{-1}$ (delivered 23 $fb^{-1}$).

## 3.3 Overview and Motivation for the LHC Detectors

The enthusiasm for the great physics potential of an LHC arose in the community at the Lausanne ECFA-CERN workshop in 1984 already mentioned before. Finding the Higgs boson, the direct experimental manifestation of the Brout-Englert-Higgs mechanism, was clearly a priority for the future of particle physics, but was also expected to be very challenging. Its mass is not predicted by the Standard Model and could have been as high as 1 TeV. This required a search over a broad range of mass, hence ideally suited at a high-energy pp collider where the energy spectrum of the constituents of the protons (quarks and gluons) allow all possible Higgs masses to be looked for at the same time. Because of the predicted low detectable cross-sections a very high-luminosity collider was mandatory.

But the Higgs search was by far not the only reason to stimulate great interest for the LHC. Already in the 1980s there were clear motivations from speculative theories predicting physics Beyond the Standard Model (BSM), the most popular one being Supersymmetry (SUSY) with its characteristic missing transverse energy signatures due to the escaping lightest neutral SUSY particle (LSP). Other hypothetical models predicted new heavy resonances, or leptoquarks (particles containing both quarks and leptons), or even substructure to quarks, and many other exotic ideas were around. It was also realized early on that the LHC would produce huge numbers of heavy flavour particles, opening a new frontier in precision flavour physics. Furthermore, unprecedented exploratory steps could be made in studying the quark-gluon plasma at high density and temperatures when colliding heavy ions, like fully ionized lead nuclei. Motivated by these physics prospects, CERN opted ultimately for an experimental programme with two very large general-purpose detectors (ATLAS and CMS), two large apparatus optimized respectively for flavour physics (LHCb) and for heavy ion collisions (ALICE), complemented later by three much smaller specialized experiments (TOTEM, total cross-section and forward physics; LHCf, measuring forward neutral particle production; MoEDAL, monopole search).

The detection of the Higgs boson played a particularly important role in the design of the **general-purpose experiments ATLAS and CMS**. In the region of low mass ($114 < m_H < 150$ GeV), the two channels considered mostly suited for unambiguous discovery were the decay to two photons and the decay to two Z bosons, each decaying in turn into $e^+e^-$ or $\mu^+\mu^-$, where one or both of the Z bosons could be virtual. As the natural width of the low-mass Higgs boson is < 10 MeV, the width of any observed peak would be entirely dominated by the instrumental mass-resolution. This meant that in designing the general-purpose detectors, considerable care was placed on the value of the magnetic field strength, on the precision tracking systems and on high-resolution em calorimeters. The high-mass region, as well as the signatures from supersymmetry, drove the need for good resolution for jets and missing transverse energy ($E_T^{miss}$), implying also almost full $4\pi$ hadronic calorimetry coverage.

The choice of the field configuration determined the overall design for these experiments. It was also well understood that to stand the best chance of making discoveries at the new energy scale of the LHC - and in the harsh conditions generated by about a billion pairs of protons interacting every second, several tens every bunch crossing - would require the invention of new technologies while at the same time pushing existing ones to their limits. In fact, a prevalent saying was "We think we know how to build a high-energy, high-luminosity hadron collider – but we don't have the technology to build a detector for it". In reality of course both turned out to be difficult and demanded technological breakthroughs. Early on it was realized that detectors will have to face eventually even luminosities beyond the initial LHC design to reach some of the ultimate physics goals. That the general-purpose experiments have worked so marvellously well since the start-up of the LHC is a testament to the difficult technology-choices made by the conceivers and the critical decisions made during the construction of these experiments. It is noteworthy that indeed the very same elements mentioned above were crucial in the discovery of a Higgs boson.

Very different challenges were faced for the two **special-purpose experiments LHCb and ALICE**, which are reflected in their very different specific designs. The only aspects they have in common is their operation at lower luminosity, typically at $2 \cdot 10^{32}$ cm$^{-2}$ s$^{-1}$ or lower, avoiding basically multiple events per bunch crossing, and their specific optimizations for particle identifications, as will be discussed later.

It is far beyond the scope of this lecture note to give a comprehensive description of these four sophisticated instruments that have been developed with very considerable R&D efforts, culminating in many large-scale prototype measurement campaigns in particle beams at CERN and other accelerator laboratories, over the 1990s and early 2000s. The construction of the various components took place over about 10 years, starting in the second half of the 1990s, in universities, national laboratories, and industries. Typically, after local initial testing, the components were sent to CERN, where after reception tests they were assembled and installed in the experimental caverns, followed by commissioning tests. During all this time, from the first Letters of Intent in 1992 to the operation phase, CERN's LHCC played an important role closely guiding and monitoring the experiments. Figure 3 shows photographs of the four detectors during their late installation phase, before the detectors were completely closed for operation.

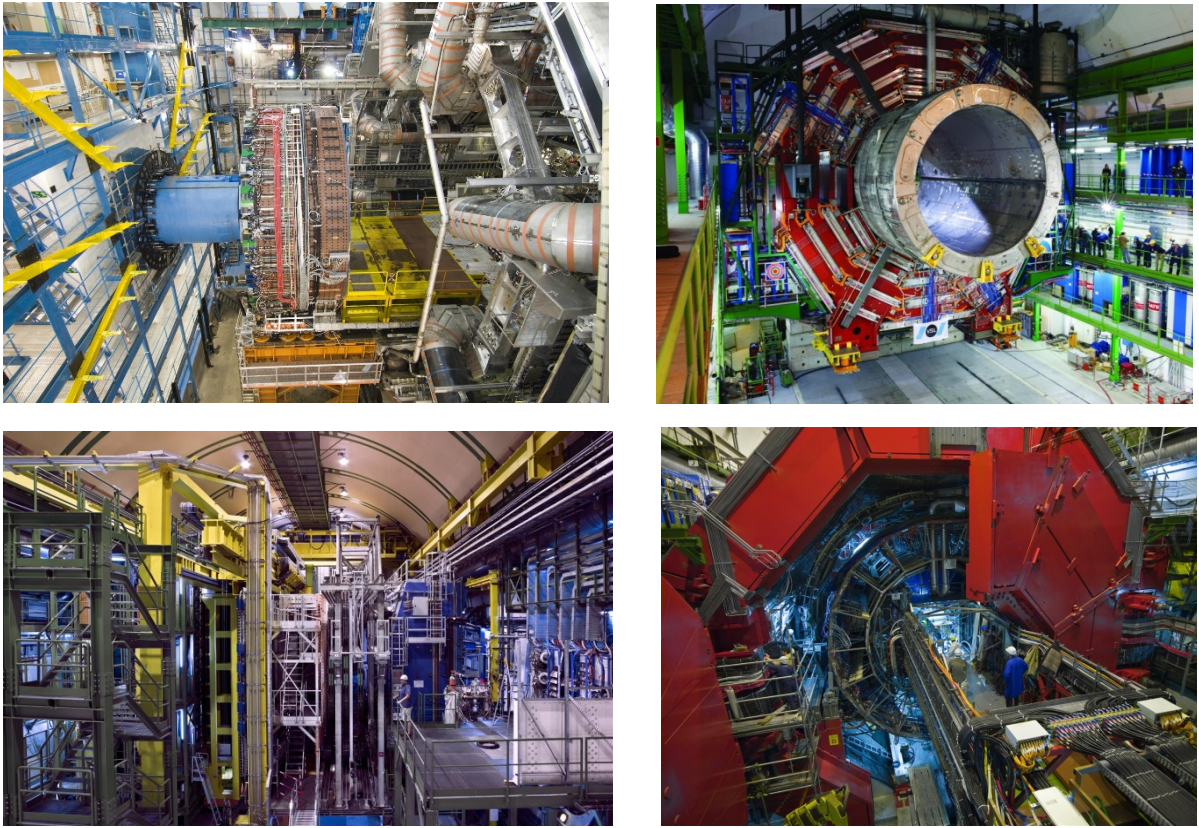

**Fig. 3:** The four LHC detectors during the installation phase. Upper row left (a) ATLAS, right (b) CMS, lower row left (c) LHCb, and right (d) ALICE.

In the following the four detectors are very briefly introduced separately. The four collaborations have published very detailed and comprehensive technical descriptions of their detectors as finally built and operated during the first years of LHC in [21-24]. The interested reader is invited to consult these major documentations for detailed information.

### 3.3.1 ATLAS [21]

The design of the ATLAS detector is based on a large toroid magnet system for the muon spectrometer complemented with a small superconducting solenoid around the inner tracking cylinder, centred at the interaction point. The novel and challenging superconducting air-core toroid magnet system, contains about 80 km of superconductor cable in eight separate barrel coils (each $25 \times 5$ m$^2$ in a 'racetrack' shape) and two matching end-cap toroids. A field of ~0.5 Tesla is generated over a large volume. The toroids are complemented with a thin solenoid (2.4 m diameter, 5.3 m length), which provides an axial magnetic field of 2 Tesla. The momentum of the muons can be precisely measured as they travel unperturbed by material for more than 5 m in the air-core toroid field. About 1200 large muon drift tube chambers of various shapes, with a total area of 5000 m$^2$, measure the impact positions with an accuracy of better than 0.1 mm. Another set of about 4200 fast chambers are used to provide a muon trigger (resistive plate chambers in the barrel, thin gap chambers in the end-caps).

In the field-free region between solenoid and toroids there is a highly granular em calorimeter using a novel Lead - liquid Argon (LAr) sampling calorimeter complemented by full-coverage hadronic sampling calorimeters. For the latter a plastic scintillator – iron sampling technique, also with a novel geometry, is used in the barrel part of the experiment, whereas LAr calorimeters cover the end-cap regions near the beam axis where particle fluxes, and thereby radiation exposures, are highest. The em and hadronic calorimeters have 200000 and 10000 cells respectively.

The reconstruction of all charged particles, including those of displaced secondary vertices, is done in the inner tracking detector, which combines highly granular pixel (50 μm x 400 μm, total 80 million channels) and microstrip (13 cm x 80 μm, total 6 million channels) silicon semiconductor sensors close to the beam axis, and a 'straw tube' gaseous detector (4 mm diameter, 350000 channels) that provides about 35 signal hits per track. The latter also helps in the identification of electrons using information from the effect of transition radiation.

Figure 3a shows one end of the cylindrical barrel detector after about 4 years of in-situ installation work in the underground cavern, 1.5 years before completion. The ends of four of the barrel toroid coils are visible, illustrating the eightfold symmetry of the structure. The relatively lightweight overall structure of the detector adds up to 7000 tonnes, spanning over a large volume of 22 m diameter with a length of 46 m.

### 3.3.2   *CMS [22]*

The CMS detector features prominently a state-of-the-art superconducting high-field solenoid of 4 Tesla. This single magnet solution serves both the high precision inner tracking chambers as well as the external muon detector, which is instrumented with large gaseous drift chambers in the barrel and cathode strip chambers in the end-caps complemented by resistive plate trigger chambers, embedded in the return yoke. This configuration allowed one to achieve a compact overall design, limiting the diameter to 15 m. The magnet yoke makes up for most of the total detector weight of 12500 tonnes.

Muon detection, and their most accurate measurement, was a priority criterion for the CMS design, followed by precision measurements for photons and electrons, achieved with a new type of radiation hard em crystal calorimeter, the largest ever built. The challenging, but very successful development and manufacture of the 75848 lead tungstate scintillating crystals, in the final set-up pointing to the interaction point, took more than a decade.

The solution to charged particle tracking was to opt for a small number of precise position measurements for each charged particle trajectory (13 layers with a position resolution of ~15 μm per measurement) leading to a large number of cells distributed inside a cylindrical volume 5.8 m long and 2.5 m in diameter: 66 million $100 \times 150$ μm$^2$ silicon pixels and 9.3 million silicon microstrips ranging from about 10 cm × 80 μm to 20 cm × 180 μm. With 198 m$^2$ of silicon detector area the CMS tracker is by far the largest silicon tracker ever built.

Finally the hadron calorimeter, comprising ~3000 small solid angle projective towers covering almost the full solid angle, is built from alternate plates of ~5 cm brass absorber and ~4 mm thick scintillator plates that sample the shower energy. The scintillation light is detected by photodetectors (hybrid photodiodes) that can operate in the strong magnetic field, as the calorimeters are placed inside the solenoid coil.

The iron yoke of the CMS detector is sectioned into five barrel wheels and three endcap disks at each end. The sectioning enabled the detector to be assembled and tested in a large surface hall while the underground cavern was being prepared. The sections, weighing between 350 and 2000 tonnes, were then lowered into the cavern (Fig. 3b) between October 2006 and January 2008, using a dedicated gantry crane system equipped with strand jacks: a pioneering use of this technology to simplify the underground assembly of large experiments.

### 3.3.3   *LHCb [23]*

The LHCb detector concept exploits the dominant rate of beauty production towards the beam directions, and for practical reasons concentrates just on one of the two sides. Away from the LHC collision region the layout therefore resembles a fixed target spectrometer, but with very unique features.

A silicon strip vertex locator (VELO) detector can be positioned during stable beams very close to the interaction region and beam line in order to measure accurately primary and secondary vertices, selecting events with b-quarks. Particle identification to cleanly identify the various final states is achieved by two ring image Cerenkov detectors (RICH), whereas the momentum measurements are based on a large-aperture warm dipole magnet generating an integrated field path of 4 Tesla-meters for trajectories going through all tracking stations (silicon strips and straw tube drift chambers) of the spectrometer. Calorimetry is provided for by sampling scintillator lead (em) and coarser scintillator iron devices. The muon detector behind the calorimeters, an absorber with iron plates of a total of 20 interaction lengths, sampled by four chamber layers, completes the LHCb spectrometer. Figure 3c shows a picture of the fully installed LHCb detector in LHC Point-8.

### 3.3.4   *ALICE [24]*

The ALICE detector has to cope with extremely high multiplicity events, characteristic of heavy ion collisions, including charged particle measurements at an as low as possible momentum threshold. Furthermore, particle identification is needed for many of the specific heavy ion physics signatures.

The ALICE Collaboration has reused the former huge L3 warm solenoid magnet providing a field of 0.5 Tesla over a large central volume. Within the magnet, with its 10000 tonnes heavy iron yoke, is located, as central tracking detector, the world's largest Time Projection Chamber (TPC) with a field cage of 5.6 m diameter and 5.4 m length, which provides precision tracking as well as particle identification by dE/dx ionization information. The innermost region, inside the TPC, facing the most extreme particle density region around the collision point, is covered by an optimized inner tracking system with silicon pixel and silicon drift detectors followed by double-sided silicon strip detectors. Several detector systems dedicated to particle identification over various limited solid angles cover the outside radius of the TPC: transition radiation and Cherenkov radiation detectors, and a state-of-the-art time-of-flight system (TOF). High resolution em calorimetry for photon measurements is implemented with lead tungstate scintillating crystals (similar characteristics as for CMS).

A muon spectrometer starting with a massive 4 m long sophisticated hadron absorber cone, and featuring a classical dipole magnet, covers on one side the solid angle from 2 to 9 degrees with respect to the beam direction. A front view of the ALICE detector nearing installation completion is shown in Fig. 3d.

### 3.4   Triggering and Computing

A particular challenge for ATLAS and CMS are the very high collision rates in the LHC, necessary for the Higgs search and studies, given the small production cross section combined with the need to investigate final states with very small branching fractions. In the first three years of operation the LHC reached a peak instantaneous luminosity of $7 \times 10^{33}$ $cm^{-2}s^{-1}$ with a 50 ns bunch spacing, which meant that the detectors had to simultaneously cope with up to ~ 50 overlapping (pile-up) events in a given bunch crossing. In the years ahead, the instantaneous luminosity is still expected to rise two- to three-fold.

It is technically not possible to store all data for all events, therefore a trigger system is used to reject large numbers of events and retain only the interesting ones from crossings with potential physics processes of interest. This is done in real time by sophisticated integrated trigger and data acquisition systems, involving custom-made fast electronics in a first stage and large computing farms in subsequent stages before the data is transferred to mass storage for further analyses. The initial data rate from up to 40 MHz bunch crossings with multiple pile-up events is thereby reduced to a few hundreds of Hz for offline analysis. A description of these systems is far beyond the scope of these lectures, see [21,22] for details, as well as [23,24] for the specific data selection and data flow challenges for LHCb and ALICE.

The LHC experiments generate huge amounts of data (tens of petabytes of data per year) requiring a fully distributed computing model. The worldwide LHC Computing Grid (wLCG) was developed to deal with this task [17,25]. With its hierarchical architecture of tiered centres it allows any user anywhere access to any data recorded or produced in the analyses steps during the lifetime of the experiments. The centre at CERN receives the raw data, carries out prompt reconstruction, almost in real time, and exports the raw and reconstructed data to the Tier-1 centres and also to Tier-2 centres for physics analysis. The Tier-0 must keep pace with the event rate of several hundred Hz of typically 1 MB of raw data per event from each experiment. The large Tier-1 centres provide also long-term storage of raw data and reconstructed data outside of CERN (as a second copy). They carry out, for example, second-pass reconstruction, when better calibration constants are available. The large number of events simulated by Monte Carlo methods and necessary for quantifying the expectations are produced mainly in Tier-2 centres.

### 3.5 Comment on Testing and Commissioning

The Individual detector components (e.g. chambers) were built and assembled in a distributed way all around the globe in the numerous participating institutes and were typically first tested at their production sites, then after delivered to CERN, and finally again after their installation in the underground caverns. The collaborations also invested enormous efforts in testing representative samples of the detectors in test beams at CERN and other accelerator laboratories around the world. These test beam campaigns not only verified that performance criteria were met over the several years of production of detector components, but were also used to prepare the calibration and alignment data for LHC operation. Very important were the so-called large combined test beam set-ups, which represented whole 'slices' of the different detector layers of the final detectors.

During the progressing installation the experiments made extensive use of the constant flow of cosmic rays impinging on Earth providing a reasonable flux of muons even at a depth of 100 m underground, typically a few hundred per second traversing the detectors. These muons were used to check the whole chain from hardware to analysis programs of the experiments, and to align the detector elements and calibrate their response prior to the pp collisions. In particular, after the LHC incident on $19^{th}$ September 2008 the experiments used the 15 months LHC down time, before the first collisions on $23^{rd}$ November 2009, to run the full detectors in very extensive cosmic ray campaigns, collecting many hundreds of millions of muon events. These runs allowed the experiments to be ready for physics operation, with already accurately pre-calibrated and pre-aligned detectors, by the time of the first pp collisions.

An excellent account of this huge and essential commissioning work has been given in previous lectures to this school series by A. Hoecker, and the reader is highly recommended to consult [26].

## 4 The Discovery of the Higgs Boson

### 4.1 Standard Model Measurements to Demonstrate the Performance

Observing, and measuring accurately, at the LHC collision energies, the production of known particles of the SM, was always considered to be a necessary stepping stone towards exploring the full potential of the LHC with its promise of new physics, firstly of the discovery of the Higgs boson. The SM processes, such as W and Z production, are often referred to as 'standard candles' for the experiments. However, there is much more value to measuring SM processes than this: never before could the SM physics be studied at a hadron collider with such sophisticated and highly accurate detectors, ultimately allowing tests of detailed predictions of the SM with unprecedented precision and minimal instrumental systematic errors.

An example of a very early measurement is shown in Fig. 4, produced only after a month or so after first high-energy collisions in spring 2010. ATLAS and CMS observed in such di-muon invariant mass distributions a 'summary' of decades of particle physics, with remarkable mass resolution.

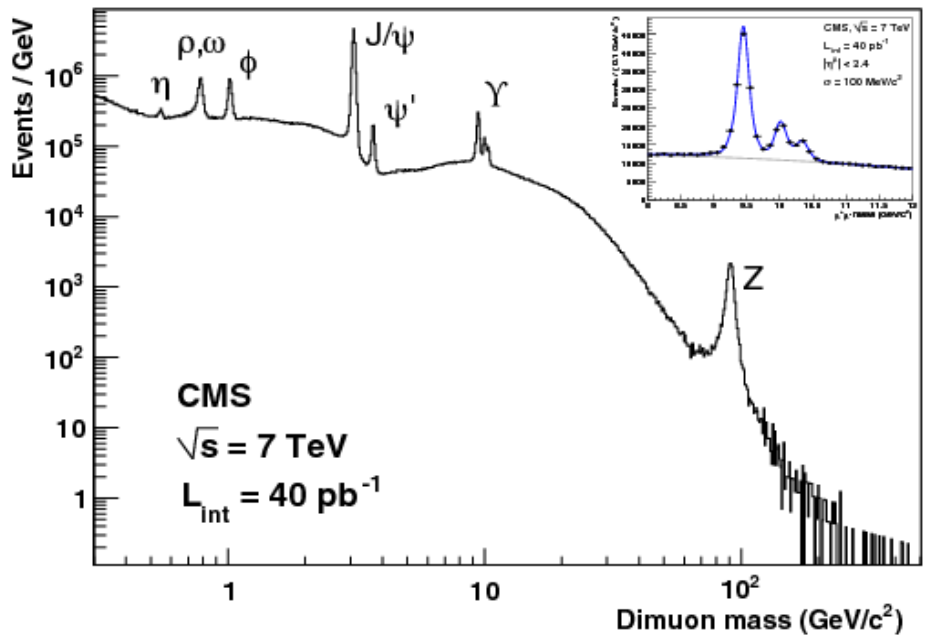

**Fig. 4:** The distribution of the invariant mass for di-muon events, shown here from CMS, displays the various well-known resonant states of the SM. The inset illustrates the excellent mass resolution for the three states of the Υ family.

In the following paragraphs a few examples will be shown from a very extensive body of publications and publically released conference contributions, which are all available at [15] for ATLAS and [16] for CMS, where detailed specific references can be found.

The charged and neutral Intermediate Vector Bosons (IVB) W and Z are the major benchmark measurements at the LHC for demonstrating the excellent detector performance, as well as for testing model predictions to a high degree of accuracy. The Z decays into electron and muon pairs can be extracted almost free of any backgrounds, as shown in Fig. 5.

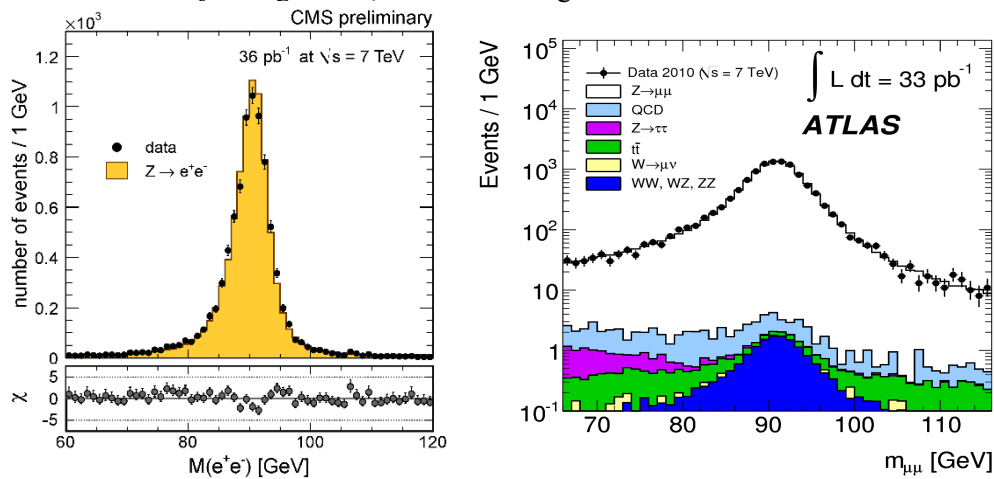

**Fig. 5:** The CMS electron-pair mass distribution on a linear (left) and the ATLAS muon-pair mass distribution on a logarithmic (right) vertical scale, in the Z mass region. The estimated small background contributions are indicated, as well as the expected signal shape from MC simulations.

The classical W decay signatures into an electron or muon and the associated neutrino are an excellent test for the $E_T^{miss}$ performance of the detector due to the undetected neutrino. $E_T^{miss}$ is inferred from the measured energy imbalance in the transverse projection of all observed signals w.r.t. the beam axis. The $E_T^{miss}$ spectrum for events with a well-identified muon candidate is shown in Fig. 6a, and shows a clear W signal over the expected SM background sources. After applying a selection of events with $E_T^{miss} > 25$ GeV only a small residual background remains present under the W signal, as indicated in the transverse mass distribution given in Fig. 6b.

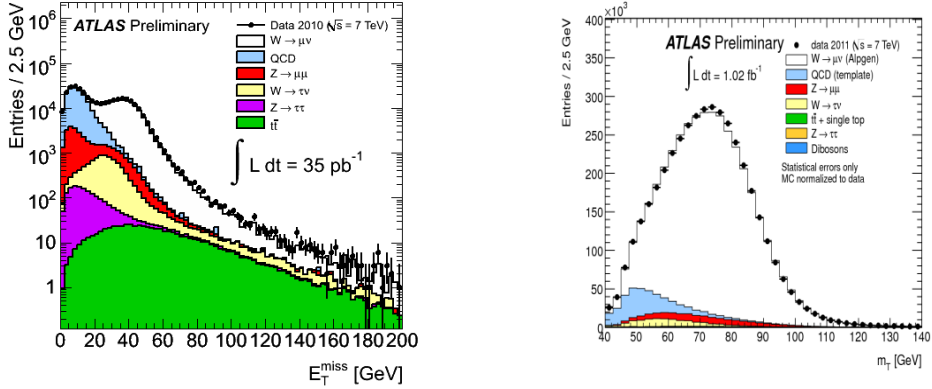

**Fig. 6:** (a, left) Missing transverse energy distribution for events with a muon candidate. (b, right) Transverse mass distribution for W to muon decays. The expected background contributions are indicated as well (examples from ATLAS).

The good agreement between the measured and expected cross-sections times the leptonic decay branching ratios (which is the expected rate for W and Z bosons to be produced and then decay to leptons) is illustrated in Fig. 7. With the presently available data samples the measurements are expected to already strongly constrain the theoretical model parameters. Figure 7a shows the cross-section measurements and predictions as a function of the collision energy, whereas in (b) the W and Z cross-section results are displayed in a 2-dimensional plot including their correlated error ellipse, and compared to predictions with various parton distribution functions (describing the quark and gluon momentum distributions inside the protons). Detailed measurements of properties for IVB production and decay at the LHC have been published already and are being refined now with the full Run-1 data samples. They include, for example, the lepton charge asymmetry measurements for W decays, which were an important signature of the electro-weak nature of the W at the time of their discovery some 30 years ago.

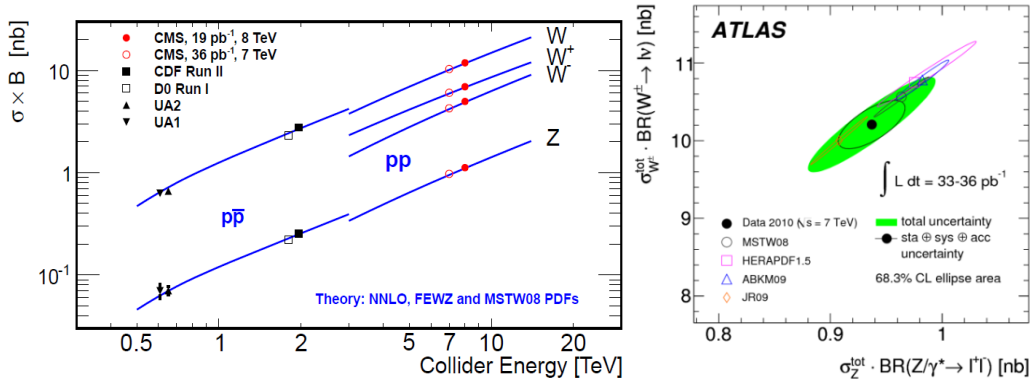

**Fig. 7:** (a, left) CMS W and Z production cross-sections times leptonic branching ratio as a function of the collision energy, showing also previous measurements at lower energy colliders. (b, right) Correlation of the measured (solid dot) leptonic W and Z cross-section from ATLAS, compared to theoretical expectations with various choices for the parton distribution functions (open symbols).

Hard collisions (characterized by having final state particles with significant transverse energy) at the LHC are dominated by the production of high transverse momentum jets, which are the collimated sprays of particles from the hadronization of the initially scattered partons (quarks, gluons) in the colliding protons. At work is the strong interaction described by Quantum Chromo Dynamics (QCD). Most commonly two jets emerge at opposite azimuth with balanced transverse momenta, from an initially lowest order parton-parton scattering process. However, higher order QCD corrections alter this picture significantly, and detailed measurements of multi-jet configurations are very important to constrain the QCD descriptions of hadronic processes.

The most impressive results at this stage are the inclusive jet and the di-jet cross-section measurements; an example from ATLAS for them is shown in Fig. 8a. These measurements cover unprecedented kinematical ranges spanning typically over jet transverse momenta from 20 GeV to 2 TeV, in many angular (pseudorapidity) bins up to $|\eta| < 4.4$ (i.e. very close to the beam axis). The cross-sections vary over these ranges by up to 12 orders of magnitude. In general the agreement with perturbative QCD calculations, including next to leading order (NLO) corrections, is well within the systematic uncertainties. This cannot be seen in Fig. 8a directly, only in ratio plots measurement/theory for a given η-interval as shown for CMS data in Fig. 8b. The systematic uncertainties in the ratios are typically less than 20%, which is a great achievement compared to previous such measurements. The systematic uncertainties on the measurements are dominated by the jet energy scale uncertainty (calibration of the detectors for the energy of jets), which, thanks to a considerable effort, has been determined to typically better than 3%.

Jets can also be produced together with W and Z bosons, so-called QCD corrections to the Intermediate Vector Boson production. Many results of these processes have been published. A good understanding of them is particularly important as they are, in many cases, a dominant source of backgrounds to the search for new particles, as well as to the measurements of top quark production discussed next.

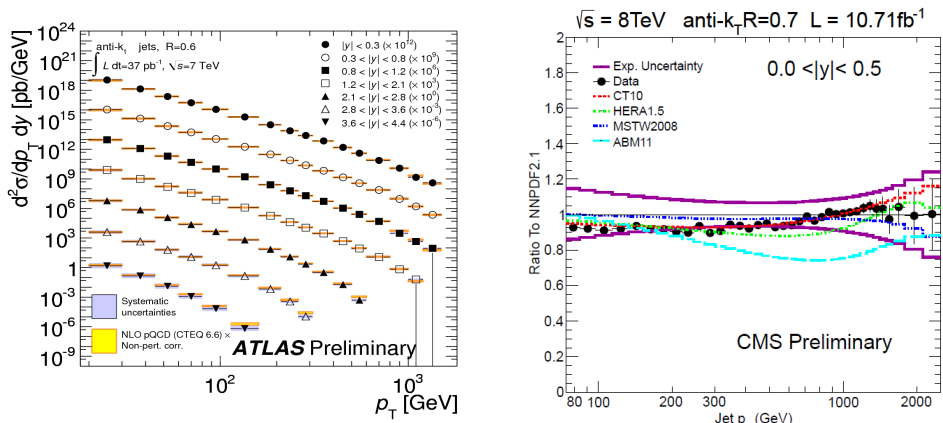

**Fig. 8:** ATLAS measurements of inclusive jet (a, left) cross-sections, and (b, right) CMS jet measurements compared to NLO perturbative QCD predictions plotted as ratio data/calculation.

The heaviest known particle in the SM is the top quark with its roughly 175 GeV mass. It decays almost exclusively into a W and a bottom quark. The measurement of top quark pair production typically requests that at least one of the W decays leptonically (also needed to trigger the events), and therefore the final states require one or two leptons (electrons or muons), $E_T^{miss}$, and jets, some of which, coming from the b-quarks, can be tagged by the displaced secondary vertices due to the finite life times of b-hadrons. Whilst it is beyond the scope of these notes to describe the sophisticated analyses employed, the message is that there are clear top pair signals in ATLAS and CMS, both in the single and two-lepton channels, when considering the correct jet topologies. The resulting cross-sections are shown in Fig. 9 which also illustrates the expected large rise of the cross-section with the collision energy increase from 2 TeV at the Tevatron to 7 TeV and 8 TeV at the LHC. Good agreement with NLO QCD calculations is seen within the present few % measurement errors. It can be mentioned that both ATLAS and CMS have also reported first single top measurements (events with just one top quark) at a rate in good agreement with QCD expectations, as well as detailed studies of top properties like its mass.

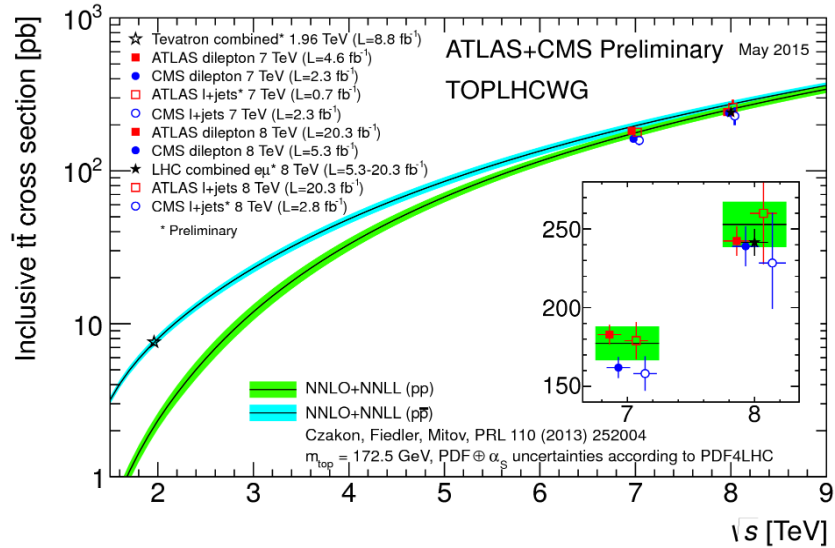

**Fig. 9:** Top pair production cross-section as a function of the collision energy, showing the Tevatron and LHC measurements.

In summary one can note that the data collected in the first three years of high-energy LHC operation have allowed ATLAS and CMS to make numerous precise measurements of SM processes, including production of bottom and top quarks, W and Z bosons, singly and in pairs. In particular very detailed measurements of QCD processes have been made. A collection of examples of such studies is shown in Fig. 10 where measurements of cross-sections for various selected electroweak and QCD processes are compared with the SM predictions. These very diverse measurements, probing cross-sections over a range of many orders of magnitude, confirm the predictions of the SM within the errors in all cases. Establishing this agreement is essential before any claims for discoveries can be made, i.e. to demonstrate on the one hand that the detector performance is well understood, and on the other hand that known SM processes are correctly observed in the experiments as they often constitute large backgrounds to signatures of new physics, such as those expected for the Higgs boson. The speed with which the wide range of measurements have shown that SM predictions for known physics have been essentially spot-on is a tribute to a large amount of work done by many particle physics theorists along with the results from the other collider experiments at LEP, Tevatron, HERA, and b-factories.

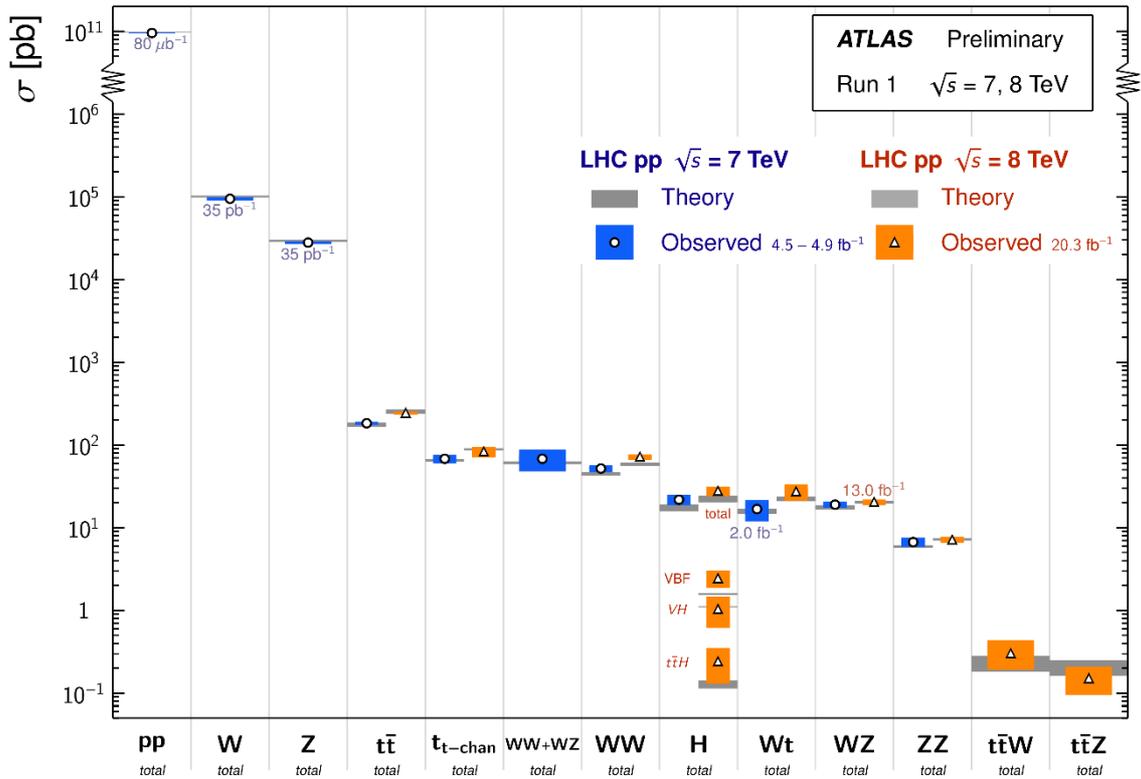

Fig. 10: A comparison of cross-section measurements for electroweak and QCD processes with theoretical predictions from the SM, shown here as example from the ATLAS experiment.

### 4.2 Discovery and Measurements of the Higgs boson

The discovery of a heavy scalar boson was announced jointly by the ATLAS and CMS Collaborations [1,2] on 4$^{th}$ July 2012 with a partial data sample corresponding to about 10 fb$^{-1}$ coming to about equal parts from running at 7 TeV collision energy in 2011 and 8 TeV in 2012 until June. The fantastic performance of the LHC during the second half of 2012 allowed the experiments to more than double their data sets. By the end of 2012 (LHC Run-1) the total amount of data that had been examined corresponded to ~5 fb$^{-1}$ at √s = 7 TeV and ~20 fb$^{-1}$ at √s = 8 TeV, equating to the examination of some 2000 trillion proton-proton collisions. Using these data first measurements of the properties of the new boson were also made. The accumulated luminosity evolution over Run-1 is illustrated in Fig. 11 for ATLAS, showing also that the experiment was very efficient in recording stably delivered luminosity as well as maintaining a high fraction (~90%) of data quality 'good for physics'.

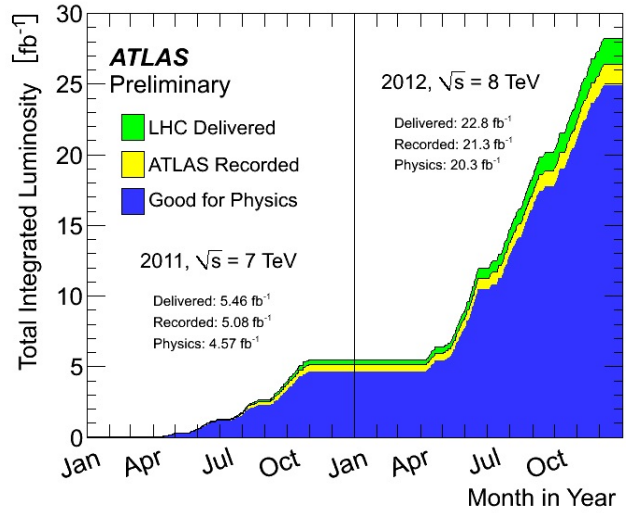

**Fig. 11:** Integrated luminosity over the high-energy running periods of Run-1 in 2011 and 2012, showing the stably delivered, recorded and finally used data sets for physics (shown is the example for ATAS, CMS is very similar).

### 4.2.1 Decays to bosons: the H→γγ, the H→ZZ→4l and H→WW→2l2ν decay modes

As examples for the full data sets, the results from the ATLAS experiment are shown for the H→γγ decay mode (Fig. 12a) and those from the CMS experiment for the H→ZZ→4$l$ mode (Fig. 12b). The signal is unmistakable and the significances are summarized in Table 2. The data show a clear excess of events above the expected background around 125 GeV. As for all the Higgs analysis figures shown in the following, the complementary data plots and results from the two experiments can be found in the detailed list of publications available from ATLAS [15] and CMS [16].

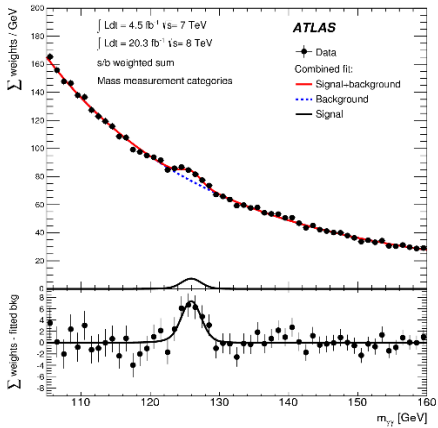
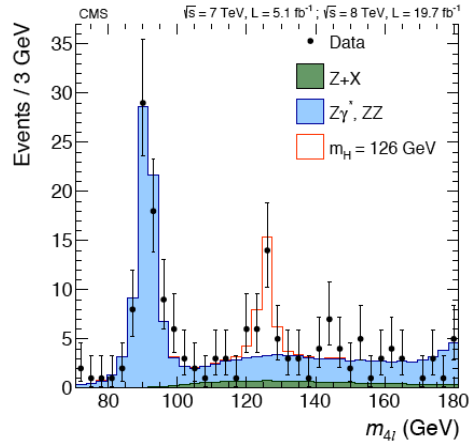

**Fig. 12a:** Invariant mass distribution of di-photon candidates. The result of a fit to the background described by a polynomial and the sum of signal components is superimposed. The bottom inset displays the residuals of the data with respect to the fitted background component.

**Fig. 12b:** The four-lepton invariant mass distribution in the CMS experiment for selected candidates relative to the background expectation. The expected signal contribution is also shown.

The search for H → WW is primarily based on the study of the final state in which both W bosons decay leptonically, resulting in a signature with two isolated, oppositely charged, high $p_T$ leptons (electrons or muons) and large $E_T^{miss}$ due to the undetected neutrinos. The signal sensitivity is improved by separating events according to lepton flavour; into $e^+e^-$, $\mu^+\mu^-$, and eμ samples, and according to jet multiplicity into 0-jet and 1-jet samples. The dominant background arises from irreducible non-resonant WW production. Any background arising from Z bosons, with same flavour but opposite sign leptons, is removed by a di-lepton mass cut $(m_Z - 15) < m_{ll} < (m_Z + 15)$ GeV.

The $m_{ll}$ distribution in the 0-jet and eµ final state is shown for CMS in Fig. 13a. The expected contribution from a SM Higgs boson with $m_H = 125$ GeV is also indicated. The transverse mass, $m_T$, distribution is shown in Fig. 13b from ATLAS, as well as the background-subtracted distribution. Both show a clear excess of events compatible with a Higgs boson with mass ~125 GeV. The observed (expected) significances of the excess with respect to the background-only hypothesis are shown in Table 2.

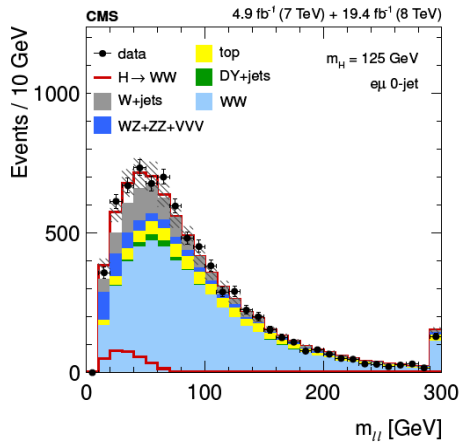 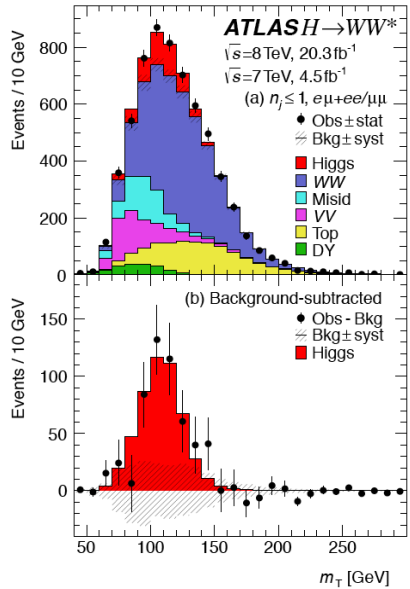

**Fig. 13a:** Distribution of dilepton mass in the 0-jet, eµ final state in CMS for a $m_H = 125$ GeV SM Higgs boson decaying via $H \rightarrow WW \rightarrow l\nu l\nu$ and for the main backgrounds

**Fig. 13b:** The transverse mass distributions for events passing the full selection of the $H \rightarrow WW \rightarrow l\nu l\nu$ analysis in ATLAS summed over all lepton flavours for final states with $N_{jet} \leq 1$. In the lower part the residuals of the data with respect to the estimated background are shown, compared to the expected $m_T$ distribution for a SM Higgs boson.

#### 4.2.2 *Decays to fermions: the $H \rightarrow \tau\tau$ and the $H \rightarrow bb$ decay modes*

It is important to establish whether this new particle also couples to fermions, and in particular to down-type fermions, since the measurements above mainly constrain the couplings to the up-type top quark. Determination of the couplings to down-type fermions requires direct measurement of the Higgs boson decays to bottom quarks and $\tau$ leptons.

The H $\rightarrow \tau\tau$ search is typically performed using the final-state signatures eµ, µµ, e$\tau_h$, µ$\tau_h$, $\tau_h\tau_h$, where electrons and muons arise from leptonic $\tau$-decays and $\tau_h$ denotes a $\tau$ lepton decaying hadronically. Each of these categories is further divided into two exclusive sub-categories based on the number and the type of the jets in the event: (i) events with one forward and one backward jet, consistent with the Vector-Boson-Fusion (VBF) topology, (ii) events with at least one high $p_T$ hadronic jet but not selected in the previous category. In each of these categories, a search is made for an excess in the reconstructed $\tau\tau$ mass distribution. The main irreducible background, Z $\rightarrow \tau\tau$ production, and the largest reducible backgrounds (W + jets, multijet production, Z → ee) are evaluated from various control samples in data.

The H→bb decay mode has by far the largest branching ratio (~54%). However since $\sigma_{bb}$ (QCD) ~ $10^7 \times \sigma$(H→ bb) the search concentrates on Higgs boson production in association with a W or Z boson using the following decay modes: W → e$\nu/\mu\nu$ and Z → ee/$\mu\mu/\nu\nu$. The Z → $\nu\nu$ decay is identified by the requirement of a large missing transverse energy. The Higgs boson candidate is reconstructed by requiring two b-tagged jets.

Evidence for a Higgs boson decaying to a $\tau\tau$ lepton pair is reported by the CMS and ATLAS Collaborations. The results are given in Table 2. The CMS results reported in Table 2 include both the H → $\tau\tau$ and H → WW contributions, considered as signal in the $\tau\tau$ decay-tag analysis. This treatment leads to an increased sensitivity to the presence of a Higgs boson that decays into both $\tau\tau$ and WW.

The CMS measurements in the H → $\tau\tau$ and VH with H → bb searches are mutually consistent, within the precision of the present data, and with the expectation for the production and decay of the SM Higgs boson. CMS has combined these two results, requiring the simultaneous analysis of the data selected by the two individual measurements. Figure 14 shows the scan of the profile likelihood as a function of the signal strength relative to the expectation for the production and decay to fermions (bb and $\tau\tau$) of a standard model Higgs boson for $m_H$ = 125 GeV. The evidence against the background-only hypothesis is found to have a maximum of 3.8$\sigma$ for $m_H$ = 125 GeV.

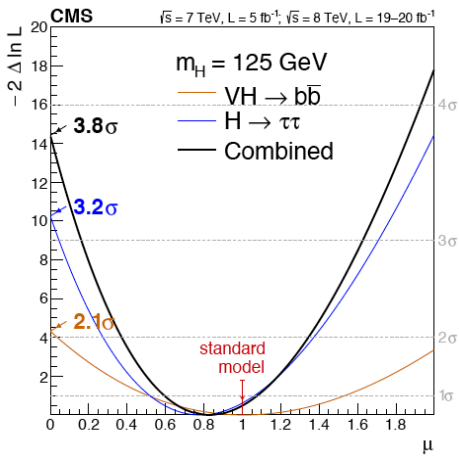 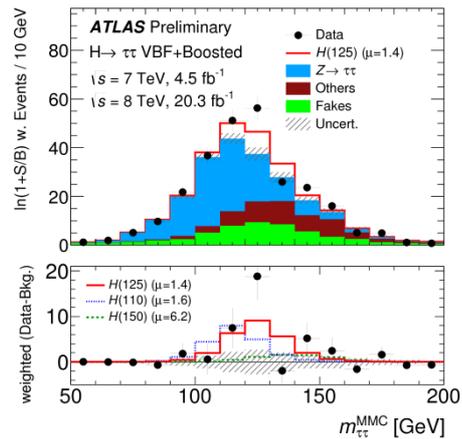

**Fig. 14:** Scan of the profile likelihood as a function of the signal strength relative to the expectation for the production and decay of a standard model Higgs boson, for $m_H$ = 125 GeV

**Fig. 15:** Observed and expected weighted di-tau mass distributions in ATLAS. The bottom panel shows the difference between weighted data events and weighted background events (points) compared to signal events yields for various masses, with signal strengths set to their best-fit values.

Figure 15 shows the observed and expected $\tau\tau$ mass distributions from the ATLAS experiment, weighing all sub-distributions in each category of each channel by the ratio between the expected signal and background yields for the respective category in a di-tau mass interval containing 68% of the signal. The plot also shows the difference between the observed data and expected background distributions, together with the expected distribution for a SM Higgs boson signal with $m_H$ = 125 GeV. The observed (expected) significance of the excess with respect to the background-only hypothesis at this mass is 4.5 (3.4) standard deviations in the ATLAS experiment.

The Tevatron experiments, CDF and D0, have also reported a combined observed significance of 3.0σ [27], where the H→ bb mode is the dominant one. All these results establish the existence of the fermionic decays of the new boson, consistent with the expectation from the SM.

*4.2.3   Higgs Boson Properties*

*4.2.3.1   The mass of the Higgs boson*

Both ATLAS and CMS experiments have separately combined their measurements of the mass of the Higgs bosons from the two channels that have the best mass resolution, namely *H→γγ* and *H→ZZ→4l*. The signal in all channels is assumed to be due to a state with a unique mass. The obtained values are from ATLAS $m_H$ = 125.36 ± 0.37(stat) ± 0.18(syst) GeV and from CMS $m_H$ = 125.02 ± 0.27(stat) ± 0.14 (syst) GeV, in excellent agreement.

*4.2.3.2   Significance of the observed excess*

Table 2 summarises the median expected and observed local significances for a SM Higgs boson mass hypothesis of 125 GeV from the individual decay modes in ATLAS and CMS. Both experiments confirm independently the discovery of a new particle with a mass near 125 GeV.

**Table 2:** The expected and observed local *p*-values in ATLAS and CMS expressed as the corresponding number of standard deviations of the observed excess from the background-only hypothesis, for $m_H$ = 125 GeV, for various decay modes.

| Experiment | ATLAS | | CMS | |
|---|---|---|---|---|
| Decay mode/combination | Expected ($\sigma$) | Observed ($\sigma$) | Expected ($\sigma$) | Observed ($\sigma$) |
| γγ | 4.6 | 5.2 | 5.3 | 5.6 |
| ZZ | 6.2 | 8.1 | 6.3 | 6.5 |
| WW | 5.8 | 6.1 | 5.4 | 4.7 |
| bb | 2.6 | 1.4 | 2.6 | 2.0 |
| ττ | 3.4 | 4.5 | 3.9 | 3.8 |
| ττ +bb  combined | - | - | 4.4 | 3.8 |

*4.2.3.3   Signal strength*

To establish whether or not the newly found state is the Higgs boson of the SM, one needs to precisely measure its other properties and attributes. Several tests of compatibility of the observed excesses with those expected from a standard model Higgs boson have been made. In one comparison labelled as the signal strength $\mu = \sigma/\sigma_{SM}$, the measured production × decay rate of the signal is compared with the SM expectation, determined for each decay mode individually and for the overall combination of all channels. A signal strength of one would be indicative of a SM Higgs boson.

Both the ATLAS and CMS experiments have measured $\mu$ values, by decay mode and by additional tags used to select preferentially events from a particular production mechanism. The best-fit value for the common signal strength $\mu$, obtained in the different sub-combinations and the overall combination of all search channels in the ATLAS and CMS experiments is shown in Fig. 16. The observed $\mu$ value is 1.00 ± 0.09 (stat) ± 0.08 (theory) for CMS for a Higgs boson mass of 125.0 GeV and 1.30 ± 0.18 in ATLAS for a Higgs boson mass of 125.5 GeV. In both the experiments the μ-values are consistent with the value expected for the SM Higgs boson ($\mu$ = 1). The Tevatron has also measured the value of this signal strength, primarily using the bb channel and find it to be 1.44±0.59 [27].

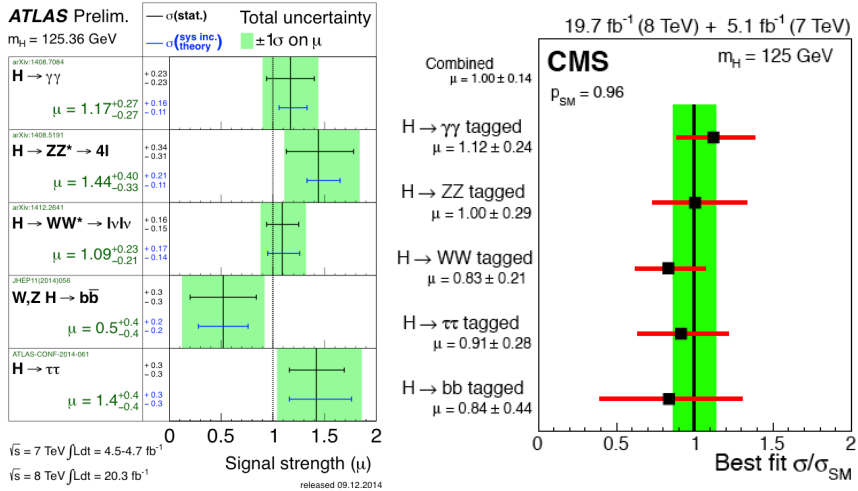

**Fig. 16:** Values of *μ* for sub-combinations by decay mode in (left) ATLAS and (right) in CMS

#### 4.2.3.4  Couplings of the Higgs boson

Figure 17 illustrates the dependence of the Higgs boson couplings on the mass of the decay particles (τ, b-quark, W, Z and t-quark). The couplings are plotted in terms of λ or √(g/2v). The line is the expectation from the SM. For the fermions, the values, λ, of the fitted Yukawa couplings *Hff* are shown, while for vector bosons the square-root of the coupling for the *HVV* vertex divided by twice the vacuum expectation value of the Higgs boson field (√(g/2v). For a Higgs boson with a mass of 125 GeV decaying to μμ CMS has found that the observed (expected) upper limit on the production rate is 7.4 (6.5 +2.8, -1.9). This corresponds to an upper limit on the branching fraction of 0.0016. The couplings are indeed proportional to mass, as expected for a SM Higgs boson, over a broad mass range, from the τ-lepton mass (about 1.8 GeV) to that of the top quark (mass about one hundred times larger).

#### 4.2.3.5  Spin and parity

Another key to the identity of the new boson is its quantum numbers amongst which is the spin-parity ($J^P$). The angular distributions of the decay particles can be used to test various spin hypotheses.

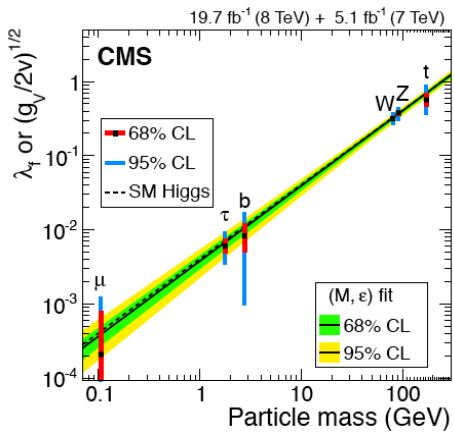

**Fig. 17:** Summary of the fits from the CMS experiment for deviations in the couplings λ or √(g/2v) as function of particle mass for a Higgs boson with a mass of 125 GeV (see text)

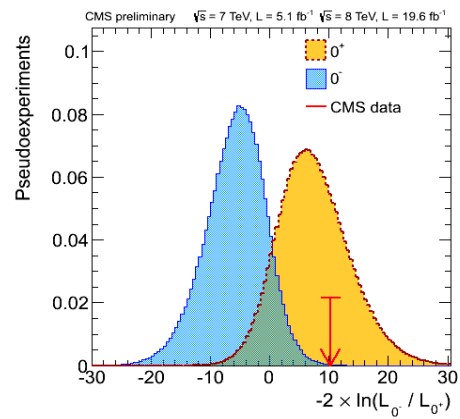

**Fig. 18:** Distribution of $q=-2ln(L_{JP}/L_{SM})$ for two signal types, $0^+$ (yellow histogram) and $0^-$ hypothesis (blue histogram) for $m_H$ = 126 GeV for a large number of generated experiments. The arrow indicates the observed value.

In the decay mode H → ZZ → 4*l* the full final state is reconstructed, including the angular variables sensitive to the spin-parity. The information from the five angles and the two di-lepton pair masses are combined to form boosted decision tree (BDT) discriminants. A decision tree is a set of cuts employed to classify events as "signal-like" or "background-like".

In the decay mode H → WW → *lνlν*, for example, in the ATLAS experiment the discriminants used in the fit are outputs of two different BDTs, trained separately against all backgrounds to identify $0^+$ and $2^+$ events, respectively. For the BDT the kinematic variables used are the transverse mass $m_T$, the azimuthal separation of the two leptons, $\Delta\phi_{ll}$, $m_{ll}$ and dilepton $p^T_{ll}$.

A first study has been presented by CMS in the ZZ → 4*l* channel with the data already disfavouring the pure pseudo-scalar hypothesis (Fig. 18). The CMS experiment has combined the ZZ → 4*l* and WW → *lνlν* spin analyses. Under the assumption that the observed boson has $J^P=0^+$, the data disfavour the hypothesis of a graviton-like boson with minimal couplings produced in gluon fusion, $J^P = 2^+$, with a CLs value of 0.60%.

ATLAS has also presented a combined study of the spin of the Higgs boson candidate using the H → γγ, H → WW → *lνlν* and H → ZZ → 4*l* decays to discriminate between the SM assignment of $J^P=0^+$ and a specific model of $J^P = 2^+$. The data strongly favour the $J^P=0^+$ hypothesis (see Fig. 19). The specific $J^P=2^+$ hypothesis is excluded with a confidence level above 99.9%, independently of the assumed contributions of gluon fusion and quark-antiquark annihilation processes in the production of the spin-2 particle.

The above-mentioned example analyses show that the spin-parity $J^P=0^+$ hypothesis is strongly favoured by both experiment, with the alternatives $J^P = 0^-, 1^+, 1^-, 2^+$ hypotheses rejected with confidence levels larger than 97.8%.

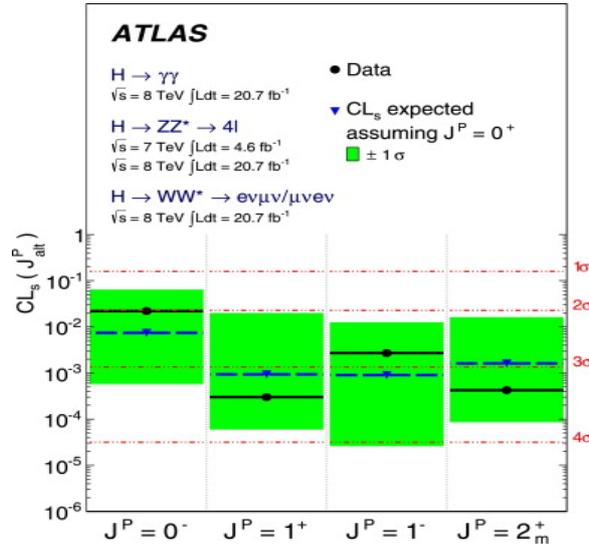

**Fig. 19:** Expected (blue triangles/dashed lines) and observed (black circles/solid lines) confidence level $CL_S$ for alternative spin–parity hypotheses assuming a $J^P=0^+$ signal. The green band represents the 68% $CL_S(J_{alt}^P)$ expected exclusion range for a signal with assumed $J^P=0^+$.

## 5    Beyond the Standard Model at the LHC

Besides the quest to elucidate the mechanism of the electro-weak symmetry breaking by searching for the Higgs boson, the major excitement for LHC comes from the great potential to explore uncharted territory of physics Beyond the Standard Model (BSM), thanks to its highest collision energy ever available in a laboratory so far. Since the beginning of the project, the search for Supersymmetry (SUSY) was a strong motivation, and besides the H boson it has been the other main benchmark physics that was guiding the detector designs. However, many other hypothetical new processes can be searched for, and indeed ATLAS and CMS have already reported in many publications a very broad spectrum of searches for BSM signatures (mass peaks for new particles or kinematical distributions with deviations from the expectations of known physics processes). No such new effect has yet been found, and all of these searches result in highly-improved, stringent exclusion limits, often well beyond the one TeV scale already. Only a few examples are mentioned below.

The most popular searches concern SUSY, which predicts additional fundamental particles. The search for SUSY is motivated in part by the prospect that the lightest stable neutral SUSY particle (LSP) could be an excellent candidate for explaining the Dark Matter (DM) in the Universe. The mysterious existence of DM was postulated by Fritz Zwicky, and rather convincingly evidenced by Vera Rubin, both astronomers, in the 1930s and 1970s respectively.

The SUSY searches at LHC are very complex as they must be sensitive to many (model-dependent) decay chains, implying a large variety of possible final state topologies. A common feature for most of them is the existence of significant missing transverse energy, $E_T^{miss}$, due to the escaping LSPs (an experimental signature similar to that of the neutrinos in the W decays). Furthermore, the SUSY signatures often include high transverse momentum jets, some tagged as b-jets for third-generation squarks as particularly motivated by naturalness arguments developed in other lectures, and leptons. The expected topologies depend not only on the model parameters, but also on the mass relations between various squarks and gluinos (the SUSY partners of the SM quarks and the gluons). A summary of 95% CL mass exclusion regions from many SUSY searches is shown in Fig. 20 from ATLAS. Very similar results are available from CMS.

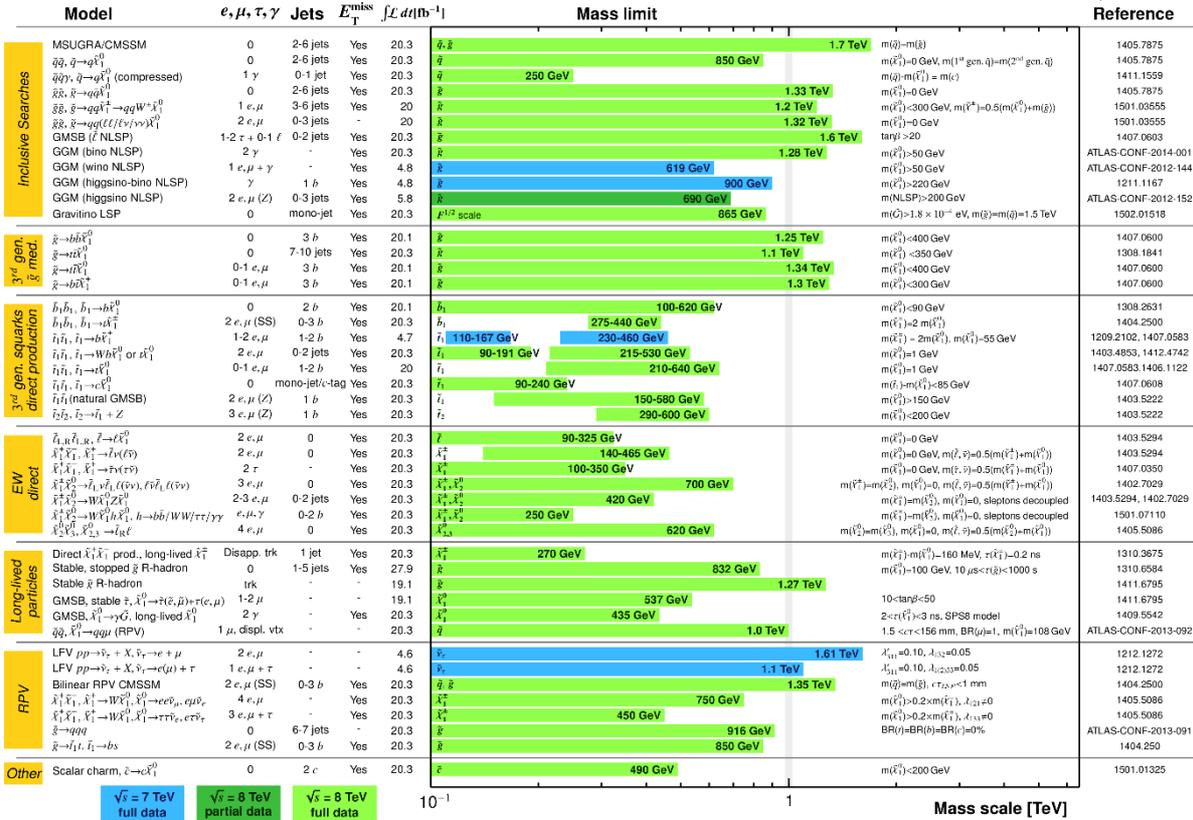

**Fig. 20:** A summary of 95% CL mass exclusion limits from many SUSY searches as obtained by ATLAS (very similar results are available from CMS as well).

Many other searches aimed at exploring BSM physics have been conducted, at this stage all without finding anywhere an excess of observed event rates over the expected backgrounds from the known SM processes. However, much more stringent limits and constraints could be established than what were available up to now. A non-exhaustive summary of 95% CL limits is displayed in Fig. 21 from CMS, with very similar results being reported by ATLAS.

Outlook

In spring 2015 the LHC will start operation again with Run-2 at the collision energy of 13 TeV and eventually 14 TeV in the coming years. Together with the increase in energy there will be also an increase in the luminosity, bringing the LHC to its full design performance (14 TeV, $2 \times 10^{34}$ cm$^{-2}$s$^{-1}$). The projected integrated luminosity by 2022 is about 300 fb$^{-1}$.

The increased energy means larger cross-sections, particularly striking for heavy objects, as can be seen in Fig. 22. It is therefore with great expectations that the experiments are looking forward to collect data in the forthcoming Run-2 and Run-3 periods, covering the initial LHC project planning.

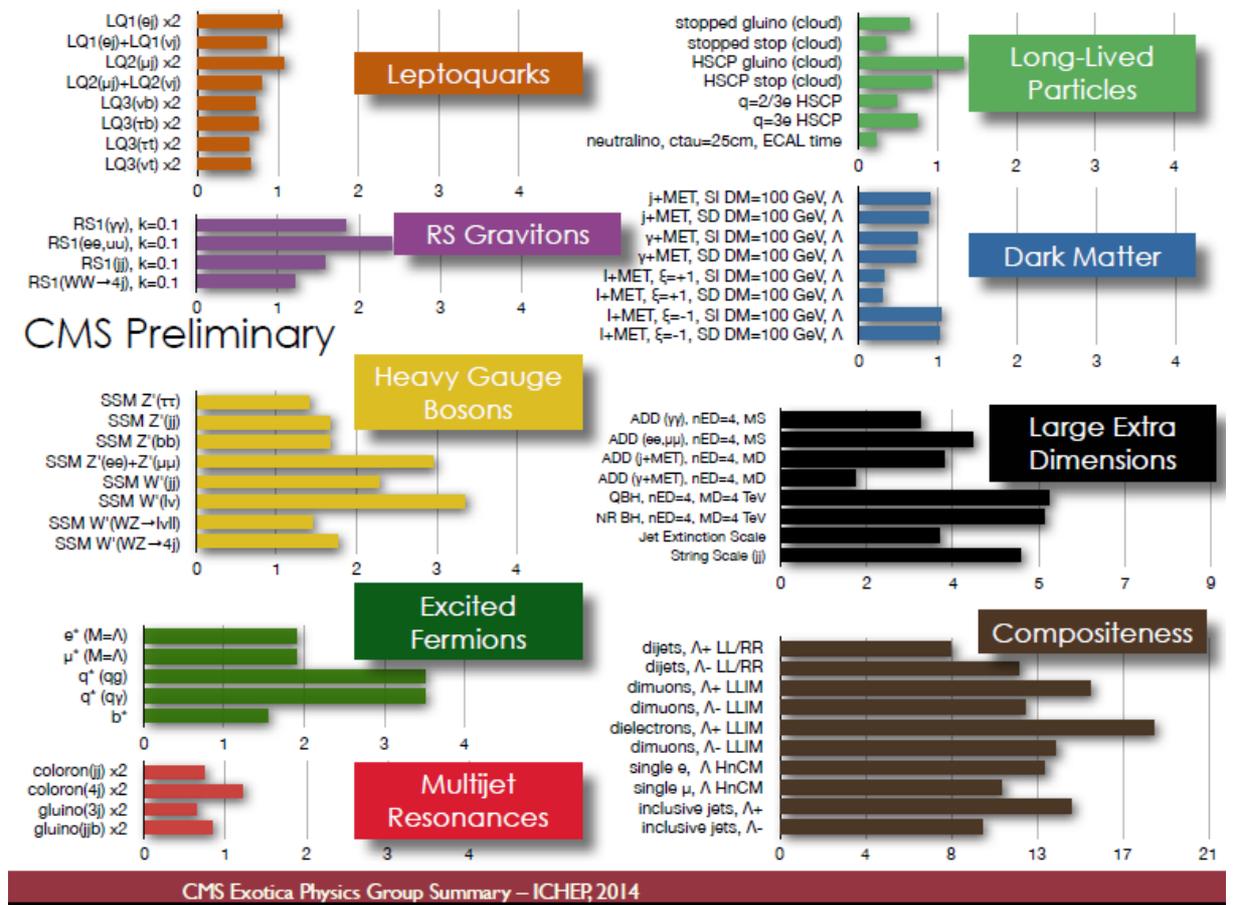

**Fig. 21:** A summary of 95% CL exclusion limits from many BSM searches other than SUSY as obtained by CMS (very similar results are available from ATLAS as well).

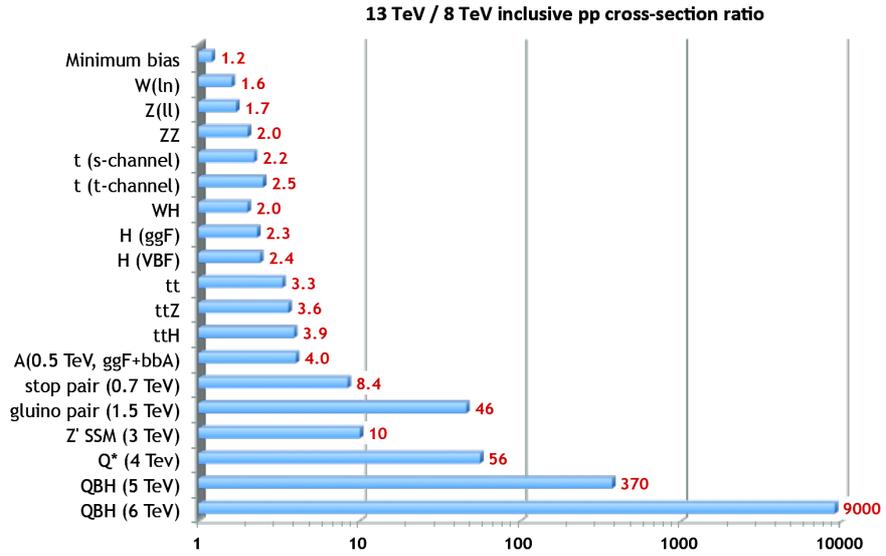

**Fig. 22:** Examples of inclusive production cross-section ratios for 13 TeV / 8 TeV

## 5.1 Prospects with the High Luminosity Upgrades (HL-LHC)

The roadmap of physics at the LHC beyond its initial design phase, with typically 300 fb$^{-1}$ integrated luminosity until the early 2020s, has dramatically changed with the discovery of the Higgs-like boson. Not only will there be the unchallenged window for directly observable hypothetical heavy mass particles, messengers of new physics beyond the Standard Model, but also a clear task to investigate in greatest details the properties of the new boson. Needless to say, this basic scenario could well be strongly enriched further if the forthcoming 14 TeV data of the current decade would reveal any new BSM physics, which would then be of course exploited best with the highest available integrated luminosity.

These prospects have strongly motivated to launch a very mayor high luminosity upgrade project planning both for the experiments and the LHC machine, called the HL-LHC, with the goal to integrate a tenfold luminosity (3000 fb-1) by the early 2030s. The importance of this future direction for particle physics has been fully recognized in the Update of the European Strategy for Particle Physics [28] where HL-LHC is singled out as first-priority in the European road map for the decades to come. A few examples are given here based on the studies in this context. The Fig. 23 illustrates the updated anticipated road map for the LHC operation for the coming decades.

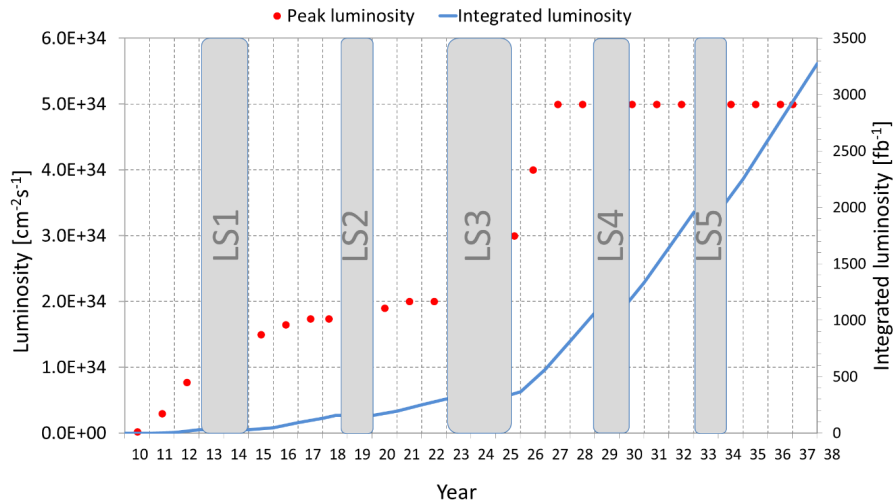

**Fig. 23:** Updated anticipated road-map for the LHC operation, with the HL-LHC starting after the Long Shut-Down number 3 (LS3) around 2025

The LHC potential for detailed studies of the electroweak symmetry breaking mechanism will be discussed first, namely the precision measurements of the Higgs couplings, the Higgs self-coupling, and vector boson scattering at high energy. In a second part a few examples of extending the reach into exploratory BSM physics will be given, including SUSY and searches for massive heavy resonances. The ATLAS and CMS Collaborations have presented a wealth of evaluations for the physics reach with the anticipated luminosity of 3000 fb$^{-1}$ for the HL-LHC era [29,30]. These estimates, given here per single experiment, are based on a very substantial simulation effort taking into account realistic pile-up conditions. Both Collaborations work on very substantial detector upgrade projects that will maintain similar detector performances as at present, and which are needed in any case for critical components to allow operation beyond the initial LHC design era. Note that similarly LHCb and ALICE have engaged into very major upgrade projects as well.

### 5.1.1 *Measurements of Higgs boson couplings*

While measurements of the Higgs boson couplings have already begun by ATLAS and CMS, these will remain a central topic within the approved LHC programme. The luminosity of the HL-LHC will provide further substantially improved statistical precision for all established channels. However, most importantly, it will also allow one to study crucial rare Higgs boson production and decay modes.

Two examples for families of channels that will only become accessible in a quantitative way with the HL-LHC are mentioned here for illustration:

- WH / ZH, H $\rightarrow$ $\gamma\gamma$ and ttH, H $\rightarrow$ $\gamma\gamma$. These channels have a low signal rate at the LHC, but one can expect to observe more than 100 events at the HL-LHC. The ttH initial state gives the cleanest signal with a signal-to-background ratio (S/B) of ~ 20%. It also provides a measurement of the top-Yukawa coupling, which is not easily accessible elsewhere. Figure 24a shows the expected signal.
- H $\rightarrow$ $\mu\mu$. The S/B of this low-rate channel is only ~ 0.2% but the narrow peak allows one to extract a more than 6 $\sigma$ significant signal for an inclusive measurement, see Fig. 24b. The exclusive ttH, H $\rightarrow$ $\mu\mu$ would yield a clean (S/B > 1) sample of 30 events providing information on both top- and $\mu$-Yukawa couplings.

An overview of the expected measurement precisions on the signal rate in each channel is given in Fig. 25 from ATLAS, but very similar results are available from CMS as well, comparing 300 and 3000 fb$^{-1}$. It should be stressed that only a limited selection of channels (initial and final states) were studied so far, and further improvements can be expected with future studies.

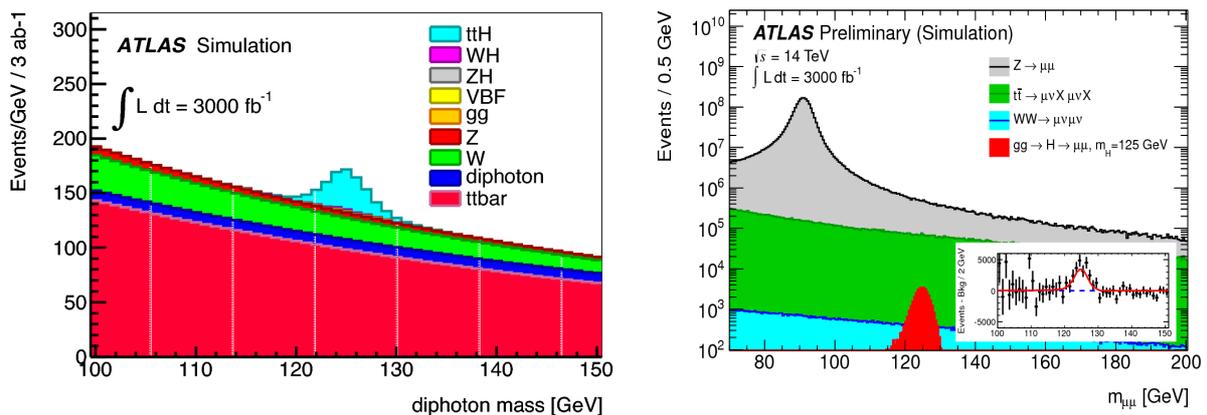

**Fig. 24:** Examples of expected invariant mass distributions for 3000 fb$^{-1}$, (a) for ttH, H $\rightarrow$ $\gamma\gamma$ selected with 1 lepton, and (b) inclusive H $\rightarrow$ $\mu\mu$.

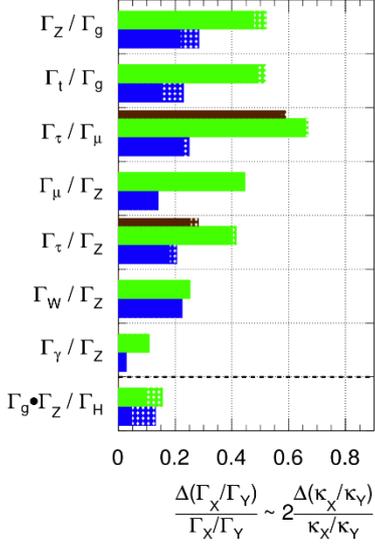 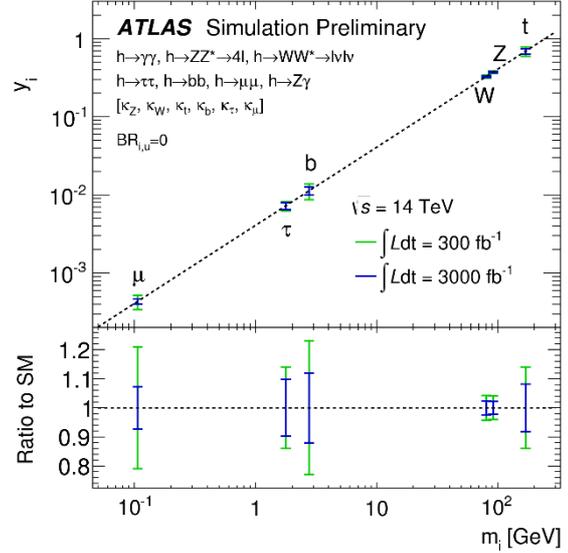

**Fig. 25:** (a) Expected precisions on ratios of Higgs boson partial widths. The bars give the expected relative uncertainty for a SM Higgs with mass 125 GeV (dashed are current theory uncertainty from QCD scale and PDFs). The thin bar for ττ show extrapolations from current analysis to 300 fb$^{-1}$, instead of the dedicated studies for VBF channels. (b) Expected precisions for the couplings (see also Fig. 17).

### 5.1.2 Observation of the Higgs self-coupling

In order to fully determine the parameters of the SM and to establish the EW symmetry breaking mechanism, the measurement of the Higgs self-coupling is important. A direct analysis of the Higgs trilinear self-coupling λ$_{HHH}$ can be done via the detection of Higgs boson pair production, through interference effects with the dominant pair production at LHC by gluon-gluon fusion. Initial sensitivity studies have been performed only on two channels so far, HH → bbγγ and bbWW, for their clean signature and high branching ratio, respectively. Only the bbγγ final state has been found to be accessible for 3000 fb$^{-1}$, yielding a 3 σ observation per experiment. Additionally, promising channels like bbττ in the final state are under investigation. The expectation is that a 30% measurement on λ$_{HHH}$ can be achieved by combining the HL-LHC measurements.

### 5.1.3 Vector boson scattering

If the new boson discovered at LHC is fully confirmed to be the SM Higgs, then unitarity of scattering amplitudes in longitudinal Vector Boson Scattering (VBS) should be preserved at high energy. It is important to confirm this prediction experimentally. It would also be important to look for new physics contributing to the regularization of the cross section or else enhancing it. For example, Technicolour or little Higgs models, postulate TeV scale resonances to become observable.

At the LHC the VBS are tagged with two forward jets on either side, the remnants of the quarks that have emitted the vector bosons involved in the scattering process. Studies of several channels have been reported for different VB decay final states for WW+2jets, WZ+2jets, and ZZ+2jets events. As an example the clean channel ZZ+2jets → 4 charged leptons + 2jets, which would allow one to fully reconstruct a hypothetical 1 TeV mass ZZ resonance peak over the SM VBS events and non-VBS di-boson background has been reported in [29].

*5.1.4 Exploratory Beyond Standard Model physics at HL-LHC*

Exploratory physics reach for BSM has always been a great motivation for the LHC, and that remains true more than ever also for the HL-LHC. Many quantitative studies exist, and have been refined now with sophisticated simulations by ATLAS and CMS with their realistic detector understanding, gained by the current LHC running in already very challenging pile-up conditions.

Considering first Supersymmetry (SUSY) searches, the new studies have confirmed that the mass reach in the generic searches for gluinos and squarks of the first two generations will be extended from typically 2.6 TeV to 3.2 TeV when adding the HL-LHC data. These results remain essentially unchanged for lightest supersymmetry particle (LSP) masses up to 1/3 of the mass of the strongly produced sparticles.

Naturalness arguments suggest the top squark to be light, preferably below 1 TeV. At 14 TeV the direct stop pair production cross section for 600 GeV (1 TeV) stops is 240 fb (10 fb). An increase in the luminosity from 300 to 3000 fb$^{-1}$ increases therefore the sensitivity significantly for heavy stop in the interesting region or, if stop candidates are found, will enable to measure their properties. As an illustrative example of a new detailed study the Fig. 26 summarizes the results in the stop-LSP plane for two decay chains. Both the 5$\sigma$ discovery range and the 95 % CL exclusion limits are shown. The cross sections for electroweak gaugino searches are small at the LHC, and the discovery potential will get strongly enhanced by the ten-fold luminosity increase. For example, the discovery potential for associated production of charginos and neutralinos extends to scenarios with chargino masses of about 800 GeV for neutralino masses below 300 GeV.

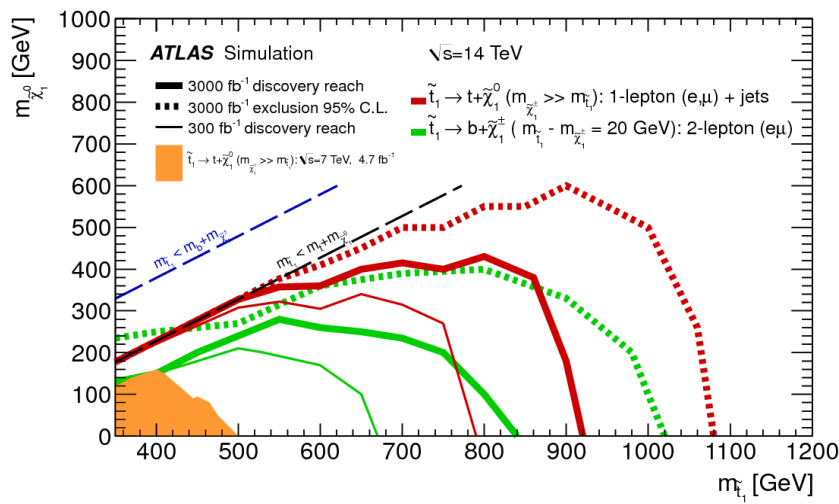

**Fig. 26:** 5$\sigma$ discovery reach and 95% CL exclusion limits in the stop-LSP mass plane for two decay channels, as indicated, for direct stop pair production.

A broad variety of resonances and other exotic signatures are sought for at the LHC. The reach for direct observations extends deep into the TeV mass scale, as a typical example one can quote the straight-forward searches for new sequential standard model like Z' decaying into charged lepton pairs. The mass reach of typically 6.5 TeV with 300 fb$^{-1}$ will increase to 7.8 TeV with 3000 fb$^{-1}$. This improved reach of about 20% is very typical for many other searches.

A notable area of exotic physics that will benefit particularly from an HL-LHC phase is the sector of final states with top quarks. Strongly and weakly produced top-antitop resonances have been studied as an interesting benchmark. For example, strongly-produced Kaluza-Klein gluons in extradimension models could result in broad top-antitop resonance signals. The mass reach for them will increase very significantly from 4.3 TeV at 300 fb$^{-1}$ to 6.7 TeV with 3000 fb$^{-1}$.

## 5.2   Prospects beyond the LHC

In the lectures an outlook for ambitious, future facilities beyond the LHC project was given. A rich variety of ambitious project dreams are pursed in the community to explore further the High Energy Frontier for many decades to come. They include hadron colliders as well as e+e- colliders (linear colliders and circular storage rings), a field too vast to describe here in this limited write-up. Studies are evolving fast, and the students are encouraged to consult the updated web information available for the CERN Future Circular Collider (FCC) studies [31], the CERN Compact Linear Collider (CLIC) studies [32], the International Linear Collider (ILC) studies [33] and the Chinese collider ring complex (CEPC-SPPC) studies [34].

With the LHC the journey into new physics territory at the high energy frontier has only just begun, and rarely before have we enjoyed such an exciting time in particle physics with great promises for discoveries. But the long LHC story, still only at the beginning of its exploitation, has also told us that timely plans and courageous decisions on a global scale have to be made by the world community of particle physicists, in order to 'plant the right seed' for the future of our field.

**Acknowledgement**

The 8th CERN Latin-American School of High-Energy Physics has been a most enjoyable time for me with great and motivating interactions with excellent students. The school was perfectly organized and I thank its organizers, Nick Ellis, Martijn Mulders and Edgar Fernando Carrera Jarrin, for having invited me to give these lectures. All the travel and stay at Ibarra has been made a pleasure by Kate Ross, whose friendly assistance is warmly acknowledged.

For this write-up I have significantly benefitted from summary articles that I had the pleasure to co-author with long-standing colleagues in the field of hadron colliders, namely Tejinder S. Virdee, Paul Grannis and Giorgio Brianti.